\documentclass[prd,twocolumn,showpacs]{revtex4}
\usepackage{graphicx, epsfig}
\usepackage{color}
\usepackage{mathrsfs}
\usepackage{bm}
\usepackage{latexsym,amssymb,amsmath,float}

\newcommand{\gapp}{\mathrel{\raise.3ex\hbox{$>$}\mkern-14mu
              \lower0.6ex\hbox{$\sim$}}}
\newcommand{\lapp}{\mathrel{\raise.3ex\hbox{$<$}\mkern-14mu
              \lower0.6ex\hbox{$\sim$}}}

\def\nn{\nonumber}
\def\x{\chi}
\def\t{\theta}
\def\p{\phi}
\def\A{{{\cal{A}}}}
\def\w{\omega}
\def\beq{\begin{eqnarray}}
\def\eeq{\end{eqnarray}}

\begin{document}
\title{Quantized Non-Abelian Monopoles on S$^3$}
\author{Irit Maor, Harsh Mathur and Tanmay Vachaspati}
\affiliation{CERCA, Department of Physics,
Case Western Reserve University, Cleveland, OH~~44106-7079}

\begin{abstract}
\noindent
A possible electric-magnetic duality suggests that the confinement
of non-Abelian electric charges manifests itself as a perturbative
quantum effect for the dual magnetic charges.
Motivated by this possibility, we study vacuum fluctuations around
a non-Abelian monopole-antimonopole pair treated as point objects
with charges $g=\pm n/2$ ($n=1,2,...$), and placed on the antipodes
of a three sphere of radius $R$. We explicitly find all the fluctuation
modes by linearizing
and solving the Yang-Mills equations about this background field on a
three sphere. We recover, generalize and extend earlier results,
including those on the stability analysis of non-Abelian magnetic
monopoles. We find that for $g \ge 1$ monopoles there is an unstable
mode that tends to squeeze magnetic flux in the angular directions.
We sum the vacuum
energy contributions of the fluctuation modes for the $g=1/2$ case
and find oscillatory dependence on the cutoff scale.
Subject to certain assumptions, we find that the contribution of the fluctuation modes to the quantum zero point energy behaves as $ -R^{-2/3}$ and hence decays more slowly than the classical $-R^{-1}$ Coulomb potential for large $R$. However, this correction to the zero point energy does not agree with the linear growth expected if the monopoles are confined.
\end{abstract}

\maketitle

\section{Introduction}
\label{introduction}

There are many facets to the investigation of magnetic monopoles.
The most straight-forward view is that magnetic monopoles are
classical solutions in certain gauge theory models
\cite{'tHooft:1974qc,Polyakov:1974ek},
and the possible realization of
these solutions leads to important ramifications for particle
physics and cosmology. Another facet is in the spirit of the
original proposal by Skyrme to understand the proton as a
classical solution \cite{Skyrme:1958vn,Skyrme:1961vr}. Then
magnetic monopoles are themselves some versions of particles, and
as suggested by the sine-Gordon and massive Thirring model
equivalence \cite{Coleman:1974bu,Mandelstam:1975hb},
magnetic monopoles are dual to ``ordinary'' electrically charged
particles whereby, in some region of parameter space, it makes
better sense to view the magnetic monopoles as the fundamental
particles and vice versa.
A tantalizing correspondence of this kind exists in grand
unified inspired models where there is a one-one correspondence
between the magnetic charges of stable monopoles and the electric
charges of known particles such as quarks and leptons
\cite{Vachaspati:1995yp,Liu:1996ea}.
Also, the duality between particles and magnetic
monopoles has been shown in certain supersymmetric theories
\cite{Seiberg:1994rs}, substantiating the early conjecture
in Ref.~\cite{Montonen:1977sn}.

If magnetic monopoles are to be viewed as magnetic versions of
electrically charged particles, perhaps it is simpler to study
certain properties of electric charges in the magnetic sector.
To follow this line of thought, we know that magnetic monopoles,
like ordinary particles, can carry non-Abelian charges. Now since
particles carrying non-Abelian charges are thought to be confined,
perhaps non-Abelian magnetic monopoles are also confined, and it
may be simpler to understand confinement by studying magnetic monopoles
instead of electric particles. This argument motivated us to study
quantum effects in the background of non-Abelian monopoles.

Confinement occurs due to non-perturbative effects in the non-Abelian
electric backgrounds of particles. The non-perturbative effects are
supposedly due to condensation of magnetically charged objects.
However, duality relates the electric and magnetic sectors with
inversely related charges. Hence, if we work in the dual sector,
the electric sources get replaced by magnetic charges (monopoles)
and, if there is confinement, it would be due to the condensation
of electric charges.  We emphasize that the exchange of
(weak, electric) and (strong, magnetic) is simply a hypothesis.
However, within this hypothesis, magnetic monopoles should get
confined by perturbative quantum effects in the electric sector.
The quantum corrections to the magnetic field of the monopole
should be calculable using a perturbative expansion in the
electric coupling.

It is to be noted that the effect we are looking
for is distinct from the confinement of magnetic monopoles
due to spontaneous symmetry breaking by the condensation of
an order parameter. We do not have extra scalar fields that
take on a vacuum expectation value to break the non-Abelian
symmetry and hence do not have any topological strings that
confine the monopoles. Instead we are hoping to see evidence
for a quantum string that confines the monopoles.
\begin{figure}
\scalebox{0.60}{\includegraphics{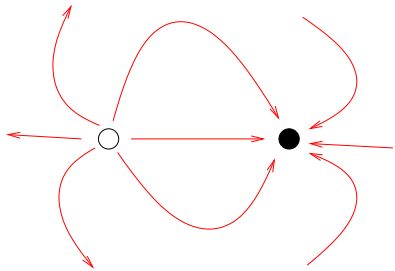}}
\caption{A classical monopole and an antimonopole in R$^3$ and their
magnetic field lines. The system has little symmetry, making it
difficult to evaluate quantum fluctuations in this background.
}
\label{mmbar}
\end{figure}

Specifically, we aim to find the one-loop quantum contribution
to the energy of a non-Abelian monopole and an anti-monopole as a
function of their separation (see Fig.~\ref{mmbar}). In this, we
have to assume a background classical field configuration for the
monopoles and this is constructed by linearly superposing the
spherically symmetric classical solutions. If the quantum
corrected energy grows faster than the separation, it would indicate
that the choice of spherically symmetric magnetic field configurations
for each monopole is not preferred since, if all the magnetic
flux is confined to a tube, the energy only grows linearly
with separation.

\begin{figure}
\scalebox{0.60}{\includegraphics{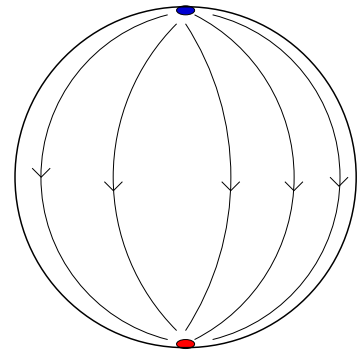}}
\caption{A monopole and an antimonopole on S$^3$ and the
classical, spherically symmetric field lines. (One dimension
of the S$^3$ has been suppressed.)
}
\label{monopolesons3}
\end{figure}

A monopole and an antimonopole in R$^3$ is a complicated
background in which to study fluctuation modes which
then have to be summed up to find the quantum corrected energy (see though \cite{backgrounds}).
Hence we have chosen to study a monopole and an antimonopole
placed at the antipodal points of an S$^3$, as shown in
Fig.~\ref{monopolesons3}. The classical magnetic field of the
monopoles is then spherically symmetric, and mitigates the difficulty
of finding the fluctuation modes. For the time being, we will
simply focus on the analysis on S$^3$, and hope that a mapping
to R$^3$ can be made at a later stage, or perhaps a similar analysis
will be found to be feasible directly in R$^3$.

Our first task is to construct non-Abelian monopoles. The simplest
model with such monopoles involves the symmetry breaking
\[
SU(3) \to [SU(2)\times U(1)]/Z_2
\]
and the fundamental monopoles have non-trivial $SU(2)$ and $U(1)$
charge. The symmetry breaking is achieved by giving a vacuum
expectation value to a field $\Phi$ transforming in the adjoint
representation of $SU(3)$. The monopoles have a regular core and
are known quite explicitly since they are basically $SU(2)$
't Hooft-Polyakov monopoles embedded within $SU(3)$
\cite{Preskill:1986kp}.

Analysis of fluctuations around monopoles arising from the $SU(3)$
model is very complicated since it involves both scalar and gauge
fields. Also, the structure of the core will be important in
determining the fluctuation modes. So we have chosen to study
non-Abelian {\em Dirac} monopoles. In this case, described more
explicitly in Sec.~\ref{background}, the monopole core is singular
but the external magnetic field for $g=1/2$ is identical to the
$SU(2)$ part of the fundamental $SU(3)$ magnetic monopole. Ignoring
the core structure does mean that we need to impose boundary
conditions at the location of the monopole by hand, an issue
that we discuss in Appendix \ref{rotational}.

In solving for the angular part of the fluctuation modes we are
aided by the early work of Wu and Yang \cite{Wu:1976ge} who found
monopole scalar harmonics, Brandt and Neri \cite{Brandt:1979kk},
who studied the stability of non-Abelian monopoles, Olsen et al
\cite{Olsen:1990jm} and Weinberg \cite{Weinberg:1993sg}, who
constructed monopole vector harmonics (also \cite{Brandt:1979kk}.
The radial problem in
R$^3$ has been set up in Ref.~\cite{Olsen:1990jm}
though with the limited goal of finding a bound
state solution. Instead our analysis requires us to evaluate
{\it all} solutions of the angular and radial problems on
S$^3$, in the background of an monopole-antimonopole pair
(and not just a single monopole).

To evaluate the one-loop quantum correction to the energy of
minimal charge monopoles with $g=1/2$ on S$^3$, we follow the procedures
developed in \cite{Dashen:1974cj} (see \cite{Rajaraman:1982is,TVbook}
for more detailed expositions). We find all the eigenfrequencies
of the fluctuation modes and then sum them up to an ultraviolet
cutoff. We do not face some of the subtleties in the quantization
of solitons since we are in a compact space and so all the
fluctuation modes are discrete. However, we treat the monopoles
as point-like ({\it i.e.} non-Abelian Dirac monopoles) and have
no control on the renormalization of its mass for example. But
we are primarily interested in the dependence of the quantum
corrections to the energy as a function of the monopole-antimonopole
separation, and assuming that the mass of the Dirac monopole has
no dependence on the size of the S$^3$,  this is given by the
dependence of the sum of zero point fluctuations on $R$, which
we can indeed calculate.

We begin by describing the background of non-Abelian monopoles on
S$^3$ in Sec.~\ref{background}. This is followed in Sec.~\ref{linearized}
by a derivation of the linearized fluctuation equations. Here we see
that the non-Abelian gauge fluctuations split up into 4 types of
excitations depending on whether they are uncharged or charged and
whether they are scalar or vector in nature. The uncharged scalar
and vector sectors do not feel the non-Abelian monopole background
but do play a role in the zero point energy, and we construct all of
them on S$^3$.
Charged scalar fluctuation modes in the background of a single monopole
in R$^3$ have been considered by Wu and Yang \cite{Wu:1976ge} leading
to the Wu-Yang spherical harmonics. Here we also solve for the radial
dependence and find the full set of charged scalar harmonics on S$^3$
in the background of the monopole-antimonopole. Similarly vector
spherical harmonics in the background of a monopole in R$^3$ have
been constructed in Refs.~\cite{Brandt:1979kk,Olsen:1990jm,Weinberg:1993sg}.
Here we find all the charged vector spherical harmonics on the
monopole-antimonopole background in S$^3$, including the radial dependence
of the eigenfunctions. In Sec.~\ref{solutions} we have summarized the
technical results for the derivation of the eigenvalues and eigenfunctions
that are shown in detail in the Appendices. For the $g=1/2$ monopole,
the sum over vacuum fluctuations is done in Sec.~\ref{quantum}. We close
the paper with discussions in Sec.~\ref{discussion}.

The evaluation of the fluctuation modes is quite technical and so it
is helpful to summarize our findings qualitatively at this stage for
the reader who may not be interested in details.
As far as the stability and possible confinement of non-Abelian monopoles
is concerned, the charged vector modes are of particular interest.
These vector modes interact
with the monopole background via their charge and also by the Zeeman
coupling of their spin to the background magnetic field. In addition,
they experience the curvature of the S$^3$. If $g \ge 1$, where
$g=eq/4\pi$ and $q$ is the magnetic monopole charge, we find that
there is an unstable mode in the spectrum of fluctuations,
implying that the spherically symmetric background is unstable. This
is consistent with the result in \cite{Brandt:1979kk} but we also
find the unstable mode explicitly and show that it tends to squeeze
the monopole flux in the angular directions.
For $g =1/2$, however, this classically unstable mode is absent. In this
case, we find the eigenvalues and eigenfunctions for all the fluctuation
modes.
We then evaluate the contribution of these quantum zero point fluctuations
to the energy of the monopole background, as compared to the zero
point energy of the trivial background, by summing up the energies of
all the individual modes. If we sum up to $\Gamma_c$ modes, we find that
this contribution to the vacuum energy oscillates as a function of $\Gamma_c$.
Averaged over the oscillations the contribution is proportional to
$-\Gamma_c^{1/3}/R$. If we impose an ultraviolet cutoff on the momentum,
this corresponds to a $-R^{-2/3}$ dependence of the vacuum energy on the
radius of the S$^3$ which is directly related to the separation of the
monopole-antimonopole pair.

Additional contributions to the vacuum energy will arise from
renormalization of the parameters, in this case the mass of the
monopoles. We have taken our monopoles to be point-like and
hence do not have control over these extra contributions.
In any case, the $-R^{-2/3}$ dependence is to be contrasted with
a $+R$ growth in the energy if the monopole and antimonopole are
confined by a flux tube. Hence the energy of the
spherically symmetric contribution does not become larger than the
energy of a flux tube for some critical $R$ and the spherically
symmetric configuration is of lower energy. So we conclude that the
$g=1/2$ monopoles on S$^3$ are not confined, at least for reasons of
vacuum energy. However, the quantum zero point energy may still
turn out to be larger than the classical $-1/R$ Coulomb energy
for monopole-antimonopole separation larger than some critical value.

\section{Monopole background}
\label{background}

We consider a Yang-Mills field on a three sphere S$^3$ of radius $R$.
The S$^3$ metric in polar coordinates ($\x$, $\t$, and $\p$) is
\beq
  ds^2 = -dt^2+ R^2 \Biggl( d\x^2 + \sin^2\x \Bigl[
            d\t^2+\sin^2\t d\p^2\Bigr]\Biggr)  &&
\eeq
The Yang-Mills equations of motion are
\beq
 &&\nabla_{\mu} F^{\mu}_{~~\alpha}+
    ie\left[A_{\mu}, F^{\mu}_{~~\alpha} \right] =0   ~,
 \label{eom0}
\eeq
where $F_{\mu\nu} = \partial_{\mu}A_{\nu}-\partial_{\nu}A_{\mu}+ie\left[A_{\mu},A_{\nu} \right]$. The curvature of space is taken into account via the covariant derivative
$\nabla_{\mu}$ which acts on vector fields as
$\nabla_{\mu}A^{\nu}=\partial_{\mu}A^{\nu}+\Gamma_{\mu\sigma}^{\nu}A^{\sigma}$.
For simplicity we will consider only $SU(2)$ Yang-Mills fields in this paper. The generalization
to $SU(N)$ is straightforward.

An exact solution to the Yang-Mills equations \eqref{eom0} is
\beq
 &&A_{\p} = q\left( \frac{1-\cos\t}{4\pi} \right) \hat\sigma^{(3)} ~, \
\ \ 0 \le \t < \pi
\label{wuyangup}
\eeq
where $\hat\sigma^{(3)}$ is the direction in group space, and $q$ is the
magnetic charge. Note that the gauge potential is not defined at $\t =\pi$.
As discussed in \cite{Wu:1976ge}, we also need to define the gauge potential
on a second coordinate patch
\beq
 &&A_{\p} = - q\left( \frac{1+\cos\t}{4\pi} \right) \hat\sigma^{(3)} ~, \
\ \ 0 < \t \le \pi
\label{wuyangdown}
\eeq
The gauge fields in the region of overlap of the two coordinate patches
are related by a gauge transformation that is well-defined provided
\beq
g= \frac{eq}{4\pi} = \frac{n}{2} \ , \ \ n=0,\pm 1, \pm 2, ...
\eeq

The background field strength can be calculated from the gauge potential
and is
\beq
 && F_{\t\p} = q\frac{\sin\t}{4\pi}\hat\sigma^{(3)} ~.
\eeq
and corresponds to the canonical monopole magnetic field
\beq
 && ({\vec B})_\chi = \frac{1}{r^2 \sin\theta} F_{\t\p}
             = \frac{q}{4\pi}\frac{1}{r^2} \hat\sigma^{(3)} ~.
\eeq
where $r \equiv \sin\chi$.

This exact solution  corresponds to a monopole at the north pole ($\x=0$)
of the S$^3$ and an antimonopole at the south pole ($\x=\pi$). The solution
is singular at the location of the poles, as is the case for Dirac
monopoles. This issue could be eliminated by considering regular
't Hooft-Polyakov monopoles \cite{'tHooft:1974qc,Polyakov:1974ek}, but it
would complicate the analysis significantly and would not change the long
distance behavior of the non-Abelian system which is our primary interest
here.

\section{Linearized Yang Mills Theory}
\label{linearized}

We now consider a linear perturbation $a_{\mu}$ around the background monopole solution $A_{\mu}$. Perturbing Eq.~\eqref{eom0} and keeping terms that are first order in $a$ gives
\beq
 &&g^{\mu\nu}\left\{ \frac{}{}
    \left( \partial_{\mu}\partial_{\nu}a_{\alpha}
    -\Gamma^{\lambda}_{\mu\nu}\partial_{\lambda}a_{\alpha} \right)
    +\Gamma^{\lambda}_{\mu\alpha}\left(
    \partial_{\lambda}a_{\nu}-\partial_{\nu}a_{\lambda}\right)\right. \nn \\
 &&~~+ie\left(\frac{}{} \left[\partial_{\mu}a_{\nu} , A_{\alpha} \right]
    +\left[\partial_{\mu}A_{\nu} , a_{\alpha} \right]
    +2\left[A_{\mu} , \partial_{\nu}a_{\alpha} \right]       \right.\nn\\
 &&~~~~ -\Gamma^{\lambda}_{\mu\nu}\left( \frac{}{}
    \left[a_{\lambda},A_{\alpha} \right]+\left[A_{\lambda},a_{\alpha} \right]
    \frac{}{}\right)  \nn \\
 &&~~~~ \left. -\Gamma^{\lambda}_{\mu\alpha}\left(\frac{}{}
    \left[a_{\nu},A_{\lambda} \right]+\left[A_{\nu},a_{\lambda} \right]
    \frac{}{}\right)
    +2\left[a_{\mu} , F_{\nu\alpha} \right]  \frac{}{}\right)   \nn\\
 &&~~+(ie)^2 \left( \frac{}{}
    \left[A_{\mu} , \left[A_{\nu} , a_{\alpha} \right] \right]
    -\left[A_{\alpha} , \left[A_{\mu} , a_{\nu} \right] \right]
    \frac{}{}\right) \left.  \frac{}{}\right\}                      \nn\\
 &&+(\partial_{\alpha}g^{\mu\nu})\left(\frac{}{}
    \partial_{\mu}a_{\nu}-\Gamma^{\lambda}_{\mu\nu}a_{\lambda}
    +ie\left[A_{\mu} , a_{\nu} \right] \frac{}{}\right) \nn \\
 && -g^{\mu\nu}(\partial_{\alpha}\Gamma^{\lambda}_{\mu\nu})a_{\lambda}  = 0 ~.
 \label{eq:linearym}
\eeq
We work in the fixed background gauge \cite{p+s},
\beq
 && g^{\mu\nu}\left(\partial_{\mu}a_{\nu}-
    \Gamma^{\lambda}_{\mu\nu}a_{\lambda}+
    ie\left[A_{\mu},a_{\nu}\right]  \right)= 0 ~.
 \label{g}
\eeq
The equations of motion can be cast into a more friendly
form by decomposing the perturbation in terms of its group components, as well as
separating the time and spatial parts of the 4-vector,
\beq
 && a_{\mu} = \sum_{k=\pm, 3} a_{\mu}^{(k)}\hat\sigma^{(k)} \nn \\
 && a_{\mu}^{(k)} =
    \left(
    a_t^{(k)}, ~ {\bf{a^{(k)}}}
    \right) ~.
\eeq
Here $k=\pm, 3$ is the group index, $\hat\sigma^{(k)}$ are the Pauli matrices
with $\hat\sigma^{(\pm)} \equiv \hat\sigma^{(1)}\pm i\hat\sigma^{(2)}$. With
this decomposition, the equations decouple into $4$ sectors: the
uncharged time component which behaves as a scalar $a_t^{(3)}$, the
scalar charged component $a_t^{(+)}$, the uncharged spatial vector
${\bf{a^{(3)}}}$, and the charged vector ${\bf{a^{(+)}}}$.
The equations of motion can then be schematically written the following way:
\beq
  \left(-R^2\partial_t^2+\Delta^{(3)}\right)a^{(3)}_t &=& 0 \label{o1} \\
  \left(-R^2\partial_t^2+{\bf{\Delta^{(3)}}}\right){\bf{a^{(3)}}} &=& 0 \\
  \left(-R^2\partial_t^2+\Delta^{(+)}\right)a^{(+)}_t &=& 0 \label{o3} \\
  \left(-R^2\partial_t^2+{\bf{\Delta^{(+)}}}\right){\bf{a^{(+)}}} &=& 0 ~.
    \label{o4}
\eeq
Here $\Delta$ and ${\bf{\Delta}}$ are the scalar and vector Laplacians
respectively, and in the $(+)$ direction the derivatives are gauged,
$\partial_i \rightarrow \partial_i-ie A_i$. The vector Laplacians take into
account the coupling to the curvature of the S$^3$ sphere, via the Ricci
scalar, and in the $(+)$ direction there is also a Zeeman-like coupling
between the background monopole and the perturbations. We have not written
the equations for  the $(-)$ scalar and vector components as they are simply
the complex conjugates of the Eqs.~\eqref{o3} and \eqref{o4} of the $(+)$
direction.

The gauge condition \eqref{g} decouples into the neutral and the charged sectors,
but it does mix between the time component and the spatial vector. It can be
schematically written as
\beq
  -R\partial_ta_{t}^{(3)}+{\bf{\nabla^{(3)}a^{(3)}}} &=& 0     \label{g3}\\
  -R\partial_ta_{t}^{(+)}+{\bf{\nabla^{(+)}a^{(+)}}} &=& 0 ~.  \label{g+}
\eeq
The explicit equations can be found in Appendix \ref{a1}. The boundary
conditions to which they are subject are discussed in Appendix \ref{rotational}.

\section{Enumeration of Eigenmodes}
\label{solutions}

Constructing the solutions to the linearized Yang-Mills equations is a
long yet subtle exercise in classical mathematical analysis. The details
are given in Appendix \ref{rotational}; here we simply enumerate the solutions
and their associated frequencies. At first we will simply list all the
solutions to the equations for the fluctuations. In Sec.~\ref{gaugecondition}
we will impose the gauge conditions and also eliminate modes that are pure gauge.
Finally we will be left with the physical modes and these are summarized
in Sec.~\ref{finalsummary}.

\subsection{Neutral scalar sector}

In this sector we compute the eigenmodes of the scalar Laplacian on
S$^3$. The modes are derived in Appendix \ref{rotational}
by separation of variables and
again in Appendix \ref{groupmethod}
by the use of group theory; the latter derivation exploits the four-dimensional
rotational symmetry of the problem and shows how the solutions
organize into multiplets that are irreducible representations of the so(4)
algebra. The resulting solutions are:
\begin{equation}
a_t^{(3)} =
Y_{JM}(\theta, \phi) \xi_{n-J, J} (\chi) e^{i \omega t}.
\label{eq:neutralscalar}
\end{equation}
Here
\begin{equation}
\xi_{n-J, J} (\chi) \equiv \sin^J (\chi) G^{J+\frac{1}{2}}_{n-J} (\cos \chi)
\label{eq:radialharmonic}
\end{equation}
where $ Y_{JM} (\theta, \phi)$ are the familiar spherical harmonics and
$G$ are the Gegenbauer polynomials \cite{morse}.
The quantum numbers span the range $n = 1, 2, 3, \ldots$  and
$J = 0, 1, 2, \ldots , n$ and $M = -J, \ldots, +J$. The energy of this
solution is given by
\begin{equation}
 \Lambda \equiv \omega R  = \sqrt{n(n+2)}
 \label{es0}
\end{equation}

\subsection{Charged scalar sector}

In this sector we must compute the eigenmodes of the scalar Laplacian on S$^3$
that has been minimally coupled to the gauge field of the monopole antimonopole
pair. The solutions are obtained by separation of variables:
\begin{equation}
a_t^{(+)} = W_{JM} (\theta, \phi) \xi_{n-J, \alpha}(\chi) e^{i \omega t}
\label{eq:chargedscalar}
\end{equation}
where
\begin{equation}
\alpha = - \frac{1}{2} + \sqrt{ \left(J + \frac{1}{2}\right)^2 - g^2 }.
\label{eq:alphadefn}
\end{equation}
$W_{JM}(\theta, \phi)$ are the Wu-Yang monopole spherical harmonics
\cite{Wu:1976ge} and $ \xi_{n-J, \alpha}$ is given by
\begin{equation}
\xi_{n-J, \alpha} \equiv \sin^{\alpha} (\chi) G^{\alpha+\frac{1}{2}}_{n-J} (\cos \chi).
\label{eq:chargedradialharmonic}
\end{equation}
The quantum numbers satisfy $ n = g, g +1, g+2, \ldots$ and $J = g, g + 1, g + 2, \ldots
n$ and $M = - J \ldots J$. The energy of this solution is given by
\begin{equation}
\Lambda = \sqrt{(n - J + \alpha)( n - J + \alpha + 2 )}
\label{es+}
\end{equation}
 Note that $\alpha = J$ for $ g = 0$
and the charged solutions reduce to the neutral ones as expected.

\subsection{Neutral vector sector}

In essence we are looking for the eigenmodes of the vector wave equation on
S$^3$ here. Since this is a vector equation we expect three sets of solutions
corresponding to different states of polarization. One set of solutions,
Eq.~(\ref{eq:neutralone}), may be constructed simply by taking the gradient
of the neutral scalar eigenmodes, Eq.~(\ref{eq:neutralscalar}). The third set,
Eq.~(\ref{eq:neutralthree}), is obtained by taking the curl of the second.
The second set of solutions, Eq.~(\ref{eq:neutraltwo}) below, are obtained by
looking for transverse solutions that have no radial $\chi$ component, as
discussed in Appendix \ref{rotational}.

\noindent
First solution:
\begin{eqnarray}
a^{(3)}_{\chi} (1) & = & Y_{JM} (\theta, \phi)  \partial_{\chi} \xi_{n-J,J}(\chi) e^{i \omega t}
\nonumber \\
a^{(3)}_{\theta} (1) & = & \partial_{\theta} Y_{JM} (\theta, \phi) \xi_{n-J,J}(\chi) e^{i \omega t}
\nonumber \\
a^{(3)}_{\phi} (1) & = & \partial_{\phi} Y_{JM}(\theta, \phi) \xi_{n-J,J}(\chi) e^{i \omega t}
\nonumber \\
\Lambda & = & \sqrt{n(n+2)}
\label{eq:neutralone}
\end{eqnarray}

\noindent
Second solution:
\begin{eqnarray}
a^{(3)}_{\chi} (2) & = & 0
\nonumber \\
a^{(3)}_{\theta} (2)  & = & \frac{1}{\sin \theta} [\partial_{\phi} Y_{JM} (\theta, \phi)]
\sin \chi \xi_{n-J, J} ( \chi ) e^{i \omega t}
\nonumber \\
a^{(3)}_{\phi} (2) & = & - \sin \theta [\partial_{\theta} Y_{JM} (\theta, \phi) ]
\sin \chi \xi_{n-J, J} (\chi) e^{i \omega t}
\nonumber \\
\Lambda & = &  n + 1
\label{eq:neutraltwo}
\end{eqnarray}

\noindent
Third solution:
\begin{eqnarray}
a^{(3)}_{\chi} (3)& = & J(J+1) Y_{JM} (\theta, \phi) \frac{1}{\sin \chi} \xi_{n-J, J} (\chi) e^{i \omega t}
\nonumber \\
a^{(3)}_{\theta} (3)& = & \partial_{\theta} Y_{JM} (\theta, \phi) \partial_{\chi}
\left[ \sin \chi \xi_{n-J, J} (\chi) \right] e^{i \omega t}
\nonumber \\
a^{(3)}_{\phi} (3)& = & \partial_{\phi} Y_{JM}(\theta, \phi) \partial_{\chi}
\left[ \sin \chi \xi_{n-J, J}(\chi) \right] e^{i \omega t}
\nonumber \\
\Lambda & = & n + 1
\label{eq:neutralthree}
\end{eqnarray}
The quantum numbers span the range $n = 1, 2, 3, \ldots$,
$J = 1, 2, \ldots, n$ and $M = - J , \ldots, J$
for the second and third solution. For the first solution
the allowed range is $n = 1, 2, 3, \ldots$, $J = 0, 1, 2, \ldots, n$
and $ M = - J, \ldots, J$;
in other words the value $J=0$ is also allowed.

\subsection{Charged vector sector}

There is a close parallel between the solutions in the neutral and charged vector
sectors. With one exception, the charged solutions can be obtained from the neutral
solutions by the replacement of
spherical harmonics $Y_{JM}$ with Wu-Yang harmonics $W_{JM}$, of
$J(J+1)$ with $J(J+1) - g^2$,
and the minimal coupling substitution $ \partial_{\phi} \rightarrow \partial_{\phi}
- i A_{\phi} \equiv D_{\phi}$
where $A_{\phi}$ is the background vector potential. The exception is that the
second solution involves an additional term proportional to $g$ as explained in
Appendix \ref{rotational}.

\noindent
First Solution:
\begin{eqnarray}
a^{(+)}_{\chi}(1) & = & W_{JM} (\theta, \phi) \partial_{\chi}
\xi_{n-J, \alpha} (\chi) e^{i \omega t}
\nonumber \\
a^{(+)}_{\theta}(1) & = & \partial_{\theta}
W_{JM} (\theta, \phi) \xi_{n-J, \alpha} (\chi) e^{i \omega t}
\nonumber \\
a^{(+)}_{\phi}(1) & = &
D_{\phi} W_{JM} (\theta, \phi)
\xi_{n-J, \alpha} (\chi) e^{i \omega t}
\nonumber \\
 \Lambda & = & \sqrt{(n - J + \alpha)(n - J + \alpha + 2)}
\label{eq:chargedone}
\end{eqnarray}

\noindent
Second Solution:
\begin{eqnarray}
a^{(+)}_{\x}(2) & = & 0
\nonumber \\
a^{(+)}_{\theta}(2) & = &
\sin \chi ~ \xi
\left( [J(J+1) - g^2] \frac{i}{\sin \theta} D_{\phi} W + g i\partial_{\theta} W \right)
\nonumber \\
a^{(+)}_{\phi}(2) & = &
\sin \chi ~ \xi
\left( - [J(J+1) - g^2] i \sin \theta \partial_{\theta} W + g D_{\phi} W \right)
\nonumber \\
\Lambda & = & n+1 - J + \alpha
\label{eq:chargetwo}
\end{eqnarray}
For brevity we have written $ \xi_{n-J, \alpha} (\chi)$ as $\xi$ and
$W_{JM}(\theta, \phi)$ as $W$.

\noindent
Third solution:
\begin{eqnarray}
a^{(+)}_{\chi} (3)& =  & \left[ J(J+1) - g^2 \right]
W_{JM} \frac{1}{\sin \chi} \xi_{n-J, \alpha} e^{i \omega t}
\nonumber \\
a^{(+)}_{\theta} (3)& = & \partial_{\theta}
W_{JM} (\theta, \phi)
\partial_{\chi}
\left[ \sin \chi \xi_{n-J, \alpha} ( \chi ) \right] e^{i \omega t}
\nonumber \\
a^{(+)}_{\phi} (3)& = & D_{\phi} W_{JM} (\theta, \phi)
\partial_{\chi}
\left[ \sin \chi \xi_{n-J, \alpha} (\chi) \right] e^{i \omega t}
\nonumber \\
\Lambda & = & n +1- J + \alpha
\label{eq:chargethree}
\end{eqnarray}
In the first and third families the quantum numbers span the range
$n = g, g+1, \ldots$, $J = g, g+1, \ldots, n$ and $M = -J, \ldots, J$.
For the second family $n = g+1, g+2, \ldots$, $J = g+1, \ldots, n$ and
$M = -J, \ldots, J$.

The solutions given here are valid in both the north pole gauge,
Eq.~(\ref{wuyangup}), and the south pole gauge, Eq.~(\ref{wuyangdown}),
if we interpret $A_{\phi}$ and $W_{JM}$ appropriately.

\subsection{Exceptional modes}

In the charged vector sector there is a new set of modes provided $g \ge 1$
and these have no counterpart in the neutral sector or for $g=1/2$.
These exceptional modes occur for $J=g-1$ and $M=-J,\dots,+J$.
The radial quantum number $n$ becomes continuous as does the frequency
spectrum. The construction of these modes is described in
Appendix \ref{rotational} where we also describe the radial
functions $\xi(\chi)$. The explicit form of the modes is:
\begin{eqnarray}
a_{\chi} & = & 0
\nonumber \\
a_{\theta} & = &
\xi \sin \chi \exp[i(M+g)\phi] (\sin \theta)^{g+M -1} (1 + \cos \theta)^{-M}
\nonumber \\
a_{\phi} & = &
i \xi \sin \chi \exp[i(M+g)\phi] (\sin \theta)^{g+M} (1 + \cos \theta)^{-M}.
\nonumber \\
\label{eq:unstable}
\end{eqnarray}
The frequencies of these solutions form a continuum,
$\Lambda^2 \in (-\infty , +\infty)$.

It may seem surprising that the spectrum is a continuum
on a finite space S$^3$. This result is an artifact of treating
the monopoles as point objects. As explained in Appendix \ref{rotational} this
leads to a singular potential in the mode equation for the
$J = g-1$ angular momentum channel; the singular potential in
turn is responsible for the continuum character of the spectrum.
If we worked with a model in which the monopoles had structure
the singularity would be softened and we would presumably
obtain a discrete spectrum. However we still expect that
there would be both bound and unbound modes with $\Lambda^2$
less than and greater than zero respectively. The precise
spectrum would depend on the assumed structure of
the monopole core. Fortunately we do not need to
determine the exact spectrum for the problems we wish to consider
in this paper.

\subsection{Zero modes}

So far we have been looking at modes which are completely regular at
the poles $\chi =0,\pi$ of the S$^3$. If we admit modes that diverge
at the poles, we can construct zero modes that can be interpreted as
translations of the monopoles.

If we rotate the monopole and antimonopole by a small amount
while keeping them antipodal the change in the field configuration
will be a zero mode---a zero frequency solution to the linearized
Yang-Mills equations. It is a solution because the original and displaced
monopole fields are both exact solutions to the full Yang-Mills equations;
therefore their difference should be a solution to the linearized equation
for small displacement. The frequency vanishes because both original
and displaced configurations have the same energy.
These zero mode solutions are constructed
explicitly in Appendix \ref{rotational}. There are three independent zero mode
solutions corresponding to the three orthogonal directions in which we can
infinitesimally displace the monopole-antimonopole pair. Explicitly these
solutions are given by:

\vspace{2mm}
\noindent
First Zero Mode:
\begin{eqnarray}
  a^{(3)}_{\x} & = & 0   \nn \\
  a^{(3)}_{\t} & = & 0  \nn \\
  a^{(3)}_{\p} & = & \frac{\cos \x}{\sin \x} (1 - \cos^2 \t).
\label{eq:zeroone}
\end{eqnarray}
\vspace{2mm}

\noindent
Second Zero Mode:
\begin{eqnarray}
  a^{(3)}_{\x} & = & \frac{1}{\sin^2 \x}\Bigl(
            \frac{1-\cos\t}{\sin\t} \Bigr)\sin \p  \nonumber \\
  a^{(3)}_{\t} & = & \frac{\cos\x}{\sin\x}\Bigl(
            \frac{\cos\t-\cos^2\t}{\sin^2\t}\Bigr)\sin\p \nn \\
  a^{(3)}_{\phi} & = & -\frac{\cos\x}{\sin\x}\Bigl(
            \frac{1-\cos\t}{\sin\t}-\sin\t\cos\t  \Bigr) \cos\p
\label{eq:zerotwo}
\end{eqnarray}

\vspace{2mm}

\noindent
Third Zero Mode:
\begin{eqnarray}
  a^{(3)}_{\x} & = & \frac{1}{\sin^2 \x}\Bigl(
            \frac{1-\cos\t}{\sin\t} \Bigr)\cos \p  \nonumber \\
  a^{(3)}_{\t} & = & \frac{\cos\x}{\sin\x}\Bigl(
            \frac{\cos\t-\cos^2\t}{\sin^2\t}\Bigr)\cos\p \nn \\
  a^{(3)}_{\phi} & = & \frac{\cos\x}{\sin\x}\Bigl(
            \frac{1-\cos\t}{\sin\t}-\sin\t\cos\t  \Bigr) \sin\p
\label{eq:zerothree}
\end{eqnarray}

The zero modes lie in the uncharged sector because they correspond to the
difference of two field configurations with the same charges.
Note that the zero modes are singular along the line $\theta = \pi$. This
is because we have worked in the north pole gauge Eq.~(\ref{wuyangup}).
If we had worked in the south pole gauge, Eq.~(\ref{wuyangdown})
the solutions would have been singular along $\theta = 0$. The zero modes
did not arise in the previous analysis of the neutral sector because we had
previously restricted attention to solutions that were globally regular.

As the three zero modes listed above are associated with displacement of
the monopole and the antimonopole, they are not present when the background
is trivial. It is worth noting that there is another solution with zero
frequency, not associated with displacement, and therefore present also
for the trivial case: this is the neutral scalar sector solution
\eqref{es0} with $n = J = M = 0$.

\subsection{Gauge condition}
\label{gaugecondition}

The gauge conditions, Eqs.~(\ref{g3}) and (\ref{g+}), do not mix the neutral
and the charged sectors but, unlike the equations of motion, do mix the
scalar and vector components. Using the solutions of the previous section,
we now find linear combinations of those solutions that satisfy the gauge
conditions. In other words, we find 4-vectors of the form
$(\alpha a^{(k)}_t, \beta_j{\bf{a^{(k)}}}(j))$,
where $\alpha$ and $\beta_j$ are constants, and $j=1,2,3$ labels the 3 vector
solutions we have found. The coefficients $\alpha$ and $\beta_j$ are determined
by inserting these linear combinations into \eqref{g3} or \eqref{g+}.

We find three solutions, $a_{\mu}^{(k)}=(i\w a_t^{(k)},~{\bf a}^{(k)}(1))$,
$(0,~{\bf a}^{(k)}(2))$, and $(0,~{\bf a}^{(k)}(3))$
\footnote{Only the first solution can be combined with the time component,
because only it has the same energy eigenvalue as the time component.}.
However, the first combination, $(i\w a_t^{(k)}, ~{\bf a}^{(k)}(1))$
is pure gauge. This can be easily seen as it is the gradient
of a scalar function. As expected, there is no longitudinal mode for a
massless gauge field. Discarding this solution, we are finally left
with $2$ physical (transverse) modes for each sector.

A similar analysis of the zero modes shows that the three solutions,
Eqs.~\eqref{eq:zeroone}-\eqref{eq:zerothree} combined with $a_t^{(3)}=0$,
satisfy the gauge condition and are physical. The last zero mode, \eqref{es0}
with $n=J=M=0$, is pure gauge for the monopole-pair background, as well as
for the trivial background.

\section{Summary of physical modes}
\label{finalsummary}

After eliminating the combinations of eigenmodes that do not satisfy
the gauge condition and also the combinations that are pure gauge,
we are left with the following physical eigenmodes and energies. \\

\noindent
{\bf Trivial Background:} In the absence of the monopole-antimonopole
pair all three directions in group space ($\pm$ and $3$) look alike,
and each of them has 2 physical eigenmodes given by Eqs.~(\ref{eq:neutraltwo})
and (\ref{eq:neutralthree}) with $a_t^k=0$ (where $k=\pm,3$). Both
eigenmodes have the same frequency,
\begin{equation}
\Lambda  =  n+1 ~,
\end{equation}
where $n = 1,2,3,\ldots$, $J=1,2,\ldots ,n$ and $M=-J,\ldots ,J$.
There aren't any physical zero modes. \\

\noindent
{\bf Monopole-antimonopole background:} The neutral sector (which is
direction (3) in group space) has 3 zero modes with $\Lambda = 0$,
and two physical eigenmodes, Eqs.~(\ref{eq:neutraltwo}) and
(\ref{eq:neutralthree}), both with energy
\begin{equation}
\Lambda  =  n+1 ~,
\end{equation}
where $n = 1,2,3,\ldots$, $J=1,2,\ldots ,n$ and $M=-J,\ldots ,J$. \\
The $(+)$ and $(-)$ directions of group space are similar, and each
has 2 physical eigenmodes given by Eqs.~(\ref{eq:chargetwo}) and
(\ref{eq:chargethree}), again with
$a_t^{(\pm)}=0$.
The energy eigenvalue for both modes is
\begin{equation}
\Lambda  =  n+1 - J + \alpha
\end{equation}
where
$n=g+1,\ldots$,  $J=g+1,\ldots ,n$ and $M=-J,\ldots, +J$ for the
physical solution corresponding to Eq.~(\ref{eq:chargetwo}), and
$n=g,g+1,\ldots$, $J=g,g+1,\ldots ,n$ and $M=-J,\ldots, +J$ for the
physical solution corresponding to Eq.~(\ref{eq:chargethree}).
The parameter $\alpha$ is defined in Eq.~(\ref{eq:alphadefn}).\\
For $g \ge 1$ there is an additional exceptional mode, given by
Eq.~(\ref{eq:unstable}).
The eigenfrequencies of these solutions form a continuum,
$\Lambda^2 \in (-\infty , +\infty)$.

\section{Classical stability}

A key consideration is whether the modes are stable. The diagnostic
for stability is whether the frequency of a mode is real or imaginary,
or equivalently, whether the frequency squared is positive or negative.

We find that the lowest frequency for $g=1/2$, which occurs in the
charged vector sector for $n=1/2$ and $J=1/2$, is real and positive:
\beq
 && \min(\Lambda) = \frac{\sqrt{3}+1}{2} = 1.37
\eeq
(see Eq.~\eqref{eq:chargetwo} or \eqref{eq:chargethree}).
Hence a monopole with $g=1/2$ is classically stable.

In the case of $g \ge 1$ monopoles there is a continuum of exceptional
modes, Eq.~(\ref{eq:unstable}), that have $\Lambda^2 < 0$ and are hence
unstable. This shows that monopoles with $ g \geq 1$ are classically
unstable, a result first obtained by Brandt and Neri \cite{Brandt:1979kk}.

To visualize the instability of $g \ge 1$ monopoles, consider the $g=1$ case.
Then we have $J=g-1=0$ and hence $M=0$, and we find
\beq
f_{\theta\phi} \equiv \partial_\theta a_\phi - \partial_\phi a_\theta
 = - i \xi \sin\chi e^{i\phi} (1 - \cos\theta )
\eeq
Therefore the unstable mode develops a radial magnetic field component
at $\theta = \pi$ but none at $\theta =0$, suggesting that the instability
of $g=1$ monopoles is toward a spherically asymmetric configuration.

\section{Quantum Effects}
\label{quantum}

We now turn to the quantum correction to the energy of a monopole
antimonopole pair. We consider only the case $ g = 1/2$ since for
higher $g$ values the configuration is classically unstable. Our
objective is to analyze the divergent zero-point energy in the
monopole background. From this divergent contribution we make
inferences about how the energy of the configuration scales with
the distance $R$ between the monopole-antimonopole pair.
A faster than linear growth would suggest confinement. On the other
hand if the energy scales more slowly or if it decreases, but slower
than $1/R$, it is still significant because it suggests that at
large distance the energy of a pair is dominated by quantum effects.

We aim to compute $E_Q$, the difference in the zero-point energy of the
monopole-antimonopole and the zero-point energy in the absence of the
monopole-antimonopole,
\begin{eqnarray}
E_Q R &=&  \sum_{\rm modes} \Lambda \biggr |_{\rm m{\bar m}}
               -  \sum_{\rm modes} \Lambda \biggr |_{\rm trivial}
\label{EQR}
\end{eqnarray}
where the ${\rm m{\bar m}}$ subscript refers to $\Lambda$ in the
background of the monopole-antimonopole pair, and the ${\rm trivial}$
subscript refers to $\Lambda$ when there are no monopoles {\it i.e.}
the trivial vacuum. Once we have $E_Q$, we wish to study its dependence on the number of summed modes $\Gamma_c$, which we also relate to the distance between
the monopoles by $\Gamma_c = P R$ where $P$ is an ultraviolet
momentum cut-off.

Ideally we would sum
over modes in the ${\rm m{\bar m}}$ sector that map onto the modes
in the trivial sector. Unfortunately it is not clear there is
such a mapping. We might consider bringing the monopole and antimonopole
together until they annihilate and watch the modes evolve through this
process. Unfortunately the modes are highly degenerate when the monopole and
antimonopole are antipodal or coincident and so it is impossible to establish a
continuity between modes at these two extreme points.
Alternatively one could imagine keeping the monopole
and antimonopole fixed and study how the modes evolve as $g$ is turned
off. The problem is that for consistency $g$ must be half-integer quantized so
we cannot continuously vary $g$. Although physically variation of $g$ is
impossible, we cannot rule out that there is
some way to establish a mapping by analytically continuing
$g$ from $1/2$ to zero.

For the present work, we have subtracted the mode energies by
arranging the modes (indexed with $k$) in ascending order in energy, $\Lambda(k+1) \geq \Lambda(k) $, and fixing the number of summed modes $\Gamma_c$,
\begin{eqnarray}
  E_Q R &=&  \sum_{k=1}^{\Gamma_c} \Lambda(k) \biggr |_{\rm m{\bar m}}
               -\sum_{k=1}^{\Gamma_c} \Lambda(k) \biggr |_{\rm trivial}
  \label{EQRi}
\end{eqnarray}

The ordering of the modes according to the energy is as follows.
In the trivial
background the mode energy only depends on the $n$ quantum number
and we simply order the modes according to
$n$. In the ${\rm m{\bar m}}$ background, the energy of the mode is
$n+1 -J+\alpha$. It is easy to see that the combination
$-J+\alpha$ lies in the interval $[-1/2,0)$ and is monotonically
increasing with increasing $J$. This implies that the $J$ dependence
of the mode energies simply gives a small splitting among different
states having the same $n$ quantum number. Therefore it is
sufficient to order the modes primarily according to increasing
$n$ and then, for the same values of $n$, according to increasing
$J$.

It is obvious that the charged sector modes are different depending on
whether they are computed about the trivial background or the background
of the monopole antimonopole pair. Thus these modes will contribute to the
$E_Q$ which is the difference in the zero point energy with the monopole
background and the zero point energy with a trivial background.
A more subtle point is that the neutral
sector modes are also different about the monopole background.
In this case there are three zero modes that are absent in the fluctuation spectrum
about the trivial background, so the first mode of the neutral sector in the different backgrounds contributes differently,
\beq
 & \Lambda^{(3)}(1) \biggr |_{\rm m{\bar m}}=0 ,~~~~~~
   \Lambda^{(3)}(1) \biggr |_{\rm trivial}=1 ~. &
\eeq
Thus the neutral modes also contribute to the difference
in zero point energies $E_Q$ that we wish to calculate.

\begin{figure}
\scalebox{0.40}{\includegraphics{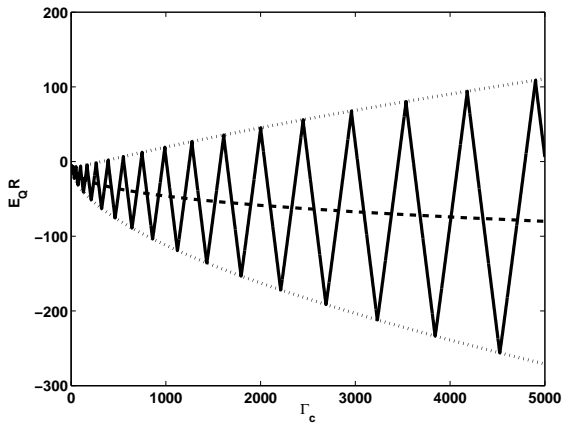}}
\caption{The energy difference $E_Q$ between the ${\rm m{\bar m}}$
vacuum and the trivial vacuum is found to oscillate as a function
of the number of modes ($\Gamma_c$) included in the energy sum
for each sector. Also drawn in are the enveloping curves for the
oscillations and the mean.
}
\label{sum}
\end{figure}

It is possible to obtain analytical bounds on $E_Q$ and these
are derived in Appendix \ref{boundsonEQ}. Here we show our
numerical results for $E_Q$, obtained after summing Eq.~(\ref{EQRi}).
Fig.~\ref{sum} shows a plot of $E_QR$ versus $\Gamma_c$, plotted in solid line.
The growing oscillatory behavior in Fig.~\ref{sum} makes it hard to
interpret the result for $E_Q$. The oscillations originate due to the
difference in the degeneracy structure of the ${\rm m\bar{m}}$ and of
the trivial case. The total angular momentum of the neutral sector is
integer, $J=1,2,..n$. For the charged sector, the total angular momentum
is half integer, $J=1/2,3/2,..n$. Ignoring the fine splitting of $J$ values
in the charged sector, the $n$th multiplet has $\sum_{J}{2(2J+1)}$ modes.
While the form is similar for both the neutral and charged sector, the
relevant $J$ span is different, resulting in multiplets of different sizes.
The first few $n$'s and their associated multiplets of both sectors are
listed in Table~\ref{table}. The result is that each sector reaches saturation
of an energy level at a different value of $\Gamma_c$. The energy sum is
dominated by the charged sector for the monopole-antimonopole background
and by the neutral sector for the trivial background, creating the zig-zag
effect of Fig.~\ref{sum}. The enveloping cone, plotted with dotted lines,
is calculated at $\Gamma_c$ values where either the neutral or the charged
sectors are saturated and jumping an energy level. Fig.~\ref{sum} also shows
the average of $E_Q$ (dashed line). As each oscillation is almost linear,
we average only the extremal points, which are on the enveloping cone.
Naming the extremal points $(\Gamma_i, E_{Qi})$, the average is calculated
according to
\beq
  \Gamma_i^{(\textrm{average})}&=& \frac{1}{4}\Bigl(
        \Gamma_{i-1}+2\Gamma_i+\Gamma_{i+1}\Bigr) \\
  E_{Qi}^{(\textrm{average})}&=& \frac{1}{4}\Bigl(
        E_{Qi-1}+2E_{Qi}+E_{Qi+1}\Bigr)
\eeq

\begin{figure}
\scalebox{0.40}{\includegraphics{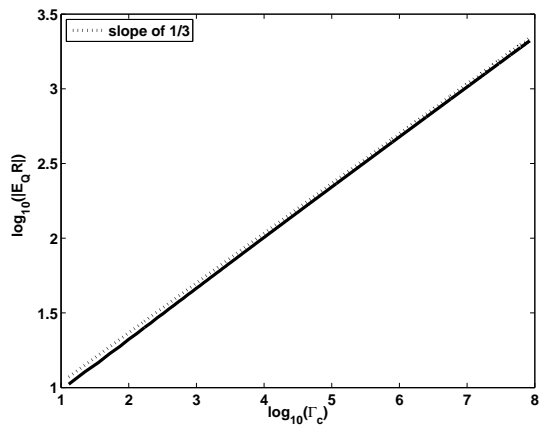}}
\caption{Smoothing the oscillatory behavior shown in Fig. \ref{sum}
reveals that on average $E_Q R$ varies as $- \Gamma_c^{1/3}$ as shown
in this plot. The smoothing was done as explained in the text.
}
\label{meanlog}
\end{figure}

\begin{figure}
\scalebox{0.40}{\includegraphics{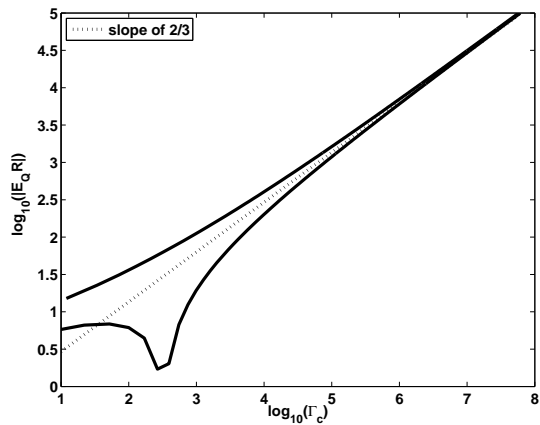}}
\caption{The amplitude of the oscillations shown in Fig.~\ref{sum}
is shown by this logarithmic plot to vary as $\Gamma_c^{2/3}$.}
\label{cone}
\end{figure}

In Fig.~\ref{meanlog} we plot average of $E_QR$ on a log-log plot and
find $\langle E_Q R \rangle \propto - \Gamma_c^{1/3}$.
We have also plotted the enveloping curves themselves on a log-log plot
in Fig.~\ref{cone}. The result is that the amplitude of the oscillations
grow in proportion to $\Gamma_c^{2/3}$; the lower enveloping curve
$\propto -\Gamma^{2/3}$, and the upper bound $\propto +\Gamma^{2/3}$.

\begin{table}
  \begin{center}
   \begin{tabular}{|c|c|}
     \hline
    {\bf{Neutral sector}} & {\bf{Charged sector}} \\
     \hline
    \begin{tabular}{c|c}
    {\bf{n}}    &   {\bf{multiplet}} \\
    \hline
    $1$ &  $6$ \\
    \hline
    $2$ &  $16$ \\
    \hline
    $3$ &  $30$ \\
    \hline
    $k$ & $2k^2+4k$
    \end{tabular}
    & \begin{tabular}{c|c}
        {\bf{n}}    &   {\bf{multiplet}} \\
    \hline
    $1/2$ & $4$ \\
    \hline
    $3/2$ & $12$ \\
    \hline
    $5/2$ & $24$ \\
     \hline
    $k+1/2$ & $2k^2 + 6k +4$
    \end{tabular} \\
%
    \hline
    \end{tabular}
    \end{center}
    \caption{The first few $n$ quantum numbers of the neutral (left)
             and charged (right) sectors, and the number of modes associated
             with that quantum number.
    \label{table}}
\end{table}

A more complete renormalization analysis would require us to evaluate
other divergent contributions to the energy of the configuration
and to extract from them a finite correction to the energy of the
configuration. We leave this problem open for later study. Such an
analysis is needed for a complete evaluation of the leading quantum
correction to the pair configuration energy.

\section{Conclusions and Discussion}
\label{discussion}

We have evaluated the complete set of vector spherical harmonics
in the background of a monopole-antimonopole pair on the antipodes
of an S$^3$. We have also solved the radial problem and thus have
a complete solution to the fluctuation problem. From our results,
we can confirm the Brandt and Neri instability for $g \ge 1$
non-Abelian monopoles \cite{Brandt:1979kk}. The unstable eigenmodes
are determined explicitly and are found to be asymmetrical in the
$\theta$ direction, suggesting an instability which might have an
interpretation in terms of flux confinement.

We show that non-Abelian monopoles with minimum charge ($g=1/2$)
do not have a classical instability. We have constructed
the complete set of fluctuation modes and their eigenvalues, and
we use these results to find the energy of the zero point fluctuations
over and above the energy of such fluctuations in the trivial
vacuum. The resulting energy contribution $E_Q$ is found to
oscillate with an amplitude that grows as $\Gamma_c^{2/3}$
where $\Gamma_c$ is the number of modes that we include
in the sum over fluctuations. The mean value of $E_Q$ however
is proportional to $-\Gamma_c^{1/3}$. If the number of modes
in the sum is limited by an ultraviolet momentum cutoff, $P$,
then $\Gamma_c = P R$ and we find that
$E_Q \propto -R^{-2/3}$. In other words, the mean contribution
of the vacuum fluctuations increases with $R$ but goes to zero
in the $R \rightarrow \infty$ limit. For confinement we would
expect $E_Q \sim R$ in the $R \rightarrow \infty$ limit and
so we conclude that our results do not provide evidence for
quantum confinement of non-Abelian magnetic monopoles, at
least on $S^3$.

Our analysis does indicate that quantum effects become important
at large $R$ because the zero point energy $\propto - R^{-2/3}$
goes to zero more slowly than the Coulomb energy $\propto -1/R$.
At some critical value of $R$, quantum effects overtake classical
effects. However, such a conclusion relies on the assumption that
a full treatment of the renormalization of the monopole mass
does not cancel out the $-R^{-2/3}$ dependence. In order to check
this assumption, we would need to quantize regular non-Abelian
monopoles {\it e.g.} $SU(3)$ monopoles as discussed in the
introduction.

A puzzling feature of our analysis is the occurrence of oscillations
in $E_Q$ as a function of the mode cutoff $\Gamma_c$. We are convinced
that these are present in our scheme of comparing energies in the
${\rm m{\bar m}}$ and trivial sectors by taking an equal number,
$\Gamma_c$, of lowest lying states in either sector. The underlying
reason is that the degeneracy structure of the modes is different
with and without monopoles. Mode degeneracies depend on the
symmetries of the system and it is clear that the ${\rm m{\bar m}}$
and trivial sectors have different symmetries under rotations.
So it is not surprising that the degeneracies in the two cases
are different. Once we accept the existence of different degeneracies,
and we compare sums with the same number of modes, then the steps
in the principal quantum number, $n$, in the sums over
modes, arise at different values of $\Gamma_c$ for the cases
with and without the monopoles. For some $\Gamma_c$ the sums with
the monopoles give a larger result and for other $\Gamma_c$ the sum
in the trivial vacuum is larger. While we understand the oscillations
mathematically, we cannot rule out the possibility that if we were to
track the flow of modes in transitioning from one sector to the other,
say by bringing the monopoles closer and letting them annihilate, it
may lead to a different subtraction scheme. We hope that future work
will shed more light on these issues.

Finally we wish to remark that the quantum interaction of non-Abelian
magnetic monopoles may be amenable to analysis by lattice methods
and this would be an alternative approach to solving the problem.

\begin{acknowledgments}
We thank Erick Weinberg for comments and suggestions. TV is grateful
to ICTP for hospitality while this work was in progress. This work
was supported by the U.S. Department of Energy and NASA at Case Western
Reserve University.
\end{acknowledgments}


\appendix
\section{Differential operators}
\label{a1}

In this appendix, we give the explicit form of Eqs.~\eqref{o1}-\eqref{g+}.

\subsection{Neutral sector operators}

Below are the equations of the uncharged
sector. The $(3)$ index of the vector components $a_i$ is suppressed.
The operator given in Eq.~\eqref{gauge3} is used in the gauge
condition, and the operator in equation \eqref{time3} is used in the
equation of motion for the scalar component.
Eqs.~\eqref{vectorx3}-\eqref{vectorp3} are the components of the vector
Laplacian, used in the vector equations of motion.
\beq
 && {\bf{\nabla^{(3)}a}} = \partial_{\x}a_{\x}+
                2\frac{\cos\x}{\sin\x}a_{\x}+ \nn \\
 && \frac{1}{\sin^2\x}\Biggl(\partial_{\t}a_{\t}+
                \frac{\cos\t}{\sin\t}a_{\t}+
                \frac{1}{\sin^2\t}\partial_{\p}a_{\p}\Biggr)
                \label{gauge3}
                \\
 && \Delta^{(3)} = \partial^2_{\x}+2\frac{\cos\x}{\sin\x}
                \partial_{\x} +
                \nn
                \\
 &&~~~~~~+ \frac{1}{\sin^2\x}\Biggl(\partial^2_{\t}+
                \frac{\cos\t}{\sin\t}\partial_{\t}+\frac{1}{\sin^2\t}
                \partial_{\p}^2\Biggr)
                \label{time3}
                \\
 &&({\bf{\Delta^{(3)} a}})_\x = \Biggl(\Delta^{(3)}-
                \frac{2}{\sin^2\x} \Biggr) a_{\x}
                - 2\frac{\cos\x}{\sin^3\x}\Biggl[
                \nn \\
 &&~~~~~~ \Biggl(
                \partial_{\t}+\frac{\cos\t}{\sin\t}\Biggr) a_{\t}+
\frac{1}{\sin^2\t}\partial_{\p}a_{\p}\Biggr]
                \label{vectorx3}
                \\
 &&({\bf{\Delta^{(3)} a}})_\t = \Biggl(\Delta^{(3)}-
                \frac{1}{\sin^2\x\sin^2\t} \Biggr) a_{\t}
                \nn \\
 &&~~~~~~+2\frac{\cos\x}{\sin\x}\Biggl(\partial_{\t}a_{\x}
                -\partial_{\x}a_{\t}  \Biggr)
                \nn \\
 &&~~~~~~-\frac{2}{\sin^2\x\sin^2\t}\frac{\cos\t}{\sin\t}
                \partial_{\p}a_{\p}
                \\
 &&({\bf{\Delta^{(3)} a}})_\p =  \Delta^{(3)} a_{\p}+
                2\frac{\cos\x}{\sin\x}\Biggl(\partial_{\p}a_{\x}-
                \partial_{\x}a_{\p}\Biggr) \nn \\
 &&~~~~~~-\frac{2}{\sin^2\x}\frac{\cos\t}{\sin\t}\Biggl( \partial_{\t}a_{\p}-
                \partial_{\p}a_{\t} \Biggr)
                \label{vectorp3}
\eeq

\subsection{Charged sector operators}

In the charged sector, derivatives are gauged, and Zeeman-like terms appear.
We use the notation of $\A_{\p}=(1-\cos\t)$, $g=eq/4\pi$, and $D_i\equiv \partial_{\p}-ig\A_{\p}$. Again, the $(+)$ index of the vector components $a_i$ is suppressed.
\beq
 && {\bf{\nabla^{(+)}a}} = \partial_{\x}a_{\x}+
                2\frac{\cos\x}{\sin\x}a_{\x}+ \nn \\
 && \frac{1}{\sin^2\x}\Biggl(\partial_{\t}a_{\t}+
                \frac{\cos\t}{\sin\t}a_{\t}+
                \frac{1}{\sin^2\t}D_{\p}a_{\p}\Biggr)
               \label{chargegauge3} \\
 && \Delta^{(+)} = \partial^2_{\x}+2\frac{\cos\x}{\sin\x}
                \partial_{\x} +
                \nn \\
 &&~~~~~~+ \frac{1}{\sin^2\x}\Biggl(\partial^2_{\t}+
                \frac{\cos\t}{\sin\t}\partial_{\t}+\frac{1}{\sin^2\t}
                D_{\p}^2\Biggr)
               \label{chargetime3} \\
 &&({\bf{\Delta^{(+)} a}})_\x = \Biggl(\Delta^{(+)}-
                \frac{2}{\sin^2\x} \Biggr) a_{\x}
                -2 \frac{\cos\x}{\sin^3\x}\Biggl[
                \nn \\
 &&~~~~~~ \Biggl(
                \partial_{\t}+\frac{\cos\t}{\sin\t}\Biggr)
                 a_{\t} +
                \frac{1}{\sin^2\t}D_{\p}a_{\p}\Biggr]
                \label{chargex3}\\
 &&({\bf{\Delta^{(+)} a}})_\t = \Biggl(\Delta^{(+)}-
                \frac{1}{\sin^2\x\sin^2\t} \Biggr) a_{\t}
                \nn \\
 &&~~~~~~+2\frac{\cos\x}{\sin\x}\Biggl(\partial_{\t}a_{\x}
                -\partial_{\x}a_{\t}  \Biggr)
                \nn \\
 &&~~~~~~-\frac{2}{\sin^2\x\sin^2\t}\Biggl( \frac{\cos\t}{\sin\t}
                D_{\p}-ig\partial_{\t}\A_{\p} \Biggr) a_{\p}
                \label{charget3}\\
 &&({\bf{\Delta a^{(+)}}})_\p =\Delta^{(+)} a_{\p}+
                2\frac{\cos\x}{\sin\x}\Biggl(D_{\p}a_{\x}-
                \partial_{\x}a_{\p}\Biggr) \nn \\
 &&~~~~~~-\frac{2}{\sin^2\x}\Biggl[ \frac{\cos\t}{\sin\t}\Biggl(
                \partial_{\t}a_{\p}-
                D_{\p}a_{\t} \biggr) \nn \\
 &&~~~~~~+ig[\partial_{\t}\A_{\p}]a_{\t} \Biggr]
 \label{chargep3}
\end{eqnarray}
Note that in the case that $g = 0$
the equations of the charged sector coincide with those of the neutral sector.

\section{Derivation of eigenmodes}
\label{rotational}

\subsection{Neutral scalar sector}

We wish to find the eigenmodes of the scalar wave equation on S$^3$.
The obvious ansatz
\begin{equation}
  a^{(3)}_t = \xi (\chi) Y_{JM} (\theta, \phi) e^{i \omega t}
\label{eq:scalarneutralansatz}
\end{equation}
leads to the eigenvalue equation
\begin{equation}
  \Biggl(\partial^2_{\chi} + 2 \frac{\cos\x}{\sin\x} \partial_{\chi} +
        \left[ \Lambda^2 - \frac{J(J+1)}{\sin^2 \chi}
        \right]\Biggl) \xi = 0.
\label{eq:neutralscalarchi}
\end{equation}
Making the substitutions $ x = \cos \chi$ and $ \xi = (1 - x^2)^{J/2} \eta$
leads to
\begin{equation}
  \Biggl( \partial_x^2 - (2 J + 3) \frac{x}{1 - x^2} \partial_x +
        \frac{ [\Lambda^2 - J(J+2)]}{1-x^2}\Biggr)  \eta = 0.
  \label{eq:gegenbauer}
\end{equation}
This is Gegenbauer's equation which has solutions regular at
$ x = \pm 1$ only for the quantized values $\Lambda^2 = n(n+2)$.
These solutions are the Gegenbauer polynomials $G^{J + \frac{1}{2}}_{n - J}(x)$
where $n - J$ must be a non-negative integer. Thus we obtain the solutions
$ \xi_{n-J, J}(\chi) Y_{JM}(\theta, \phi) \exp(i \omega t)$
where $\xi_{n-J, J}$ is given by Eq.~\eqref{eq:radialharmonic} in the
text. The quantum numbers are restricted to $n = 0, 1, \ldots$
and $J = 0, 1, \ldots n$ and $ M = - J, \ldots, J$.

As noted above the boundary conditions we impose are that the solution should
be regular at $\chi = 0$ and $\chi = \pi$. The physical basis for these boundary
conditions is that in the vicinity of the monopole, $\x=0$, the background
field strength diverges as $(F^{\mu\nu}F_{\mu\nu})^{1/2}\propto\sin^{-2}\x$,
and there is a  similar divergence in the vicinity of the antimonopole,
$\x=\pi$. In order for the perturbative expansion used in Eq.~(\ref{eq:linearym})
to be valid, the divergence of the field perturbation $f_{\mu\nu}$ cannot
be stronger than the background divergence. If the divergence is weaker than
the background then
\beq
  \lim_{\x\rightarrow 0}\Biggl(\frac{f^{\mu\nu}f_{\mu\nu}}
          {F^{\mu\nu}F_{\mu\nu}}\Biggr)\rightarrow 0 ~, &&
  \label{boundary}
\eeq
or if it is as strong as the background solution, then
\beq
  \lim_{\x\rightarrow 0}\Biggl(\frac{f^{\mu\nu}f_{\mu\nu}}
        {F^{\mu\nu}F_{\mu\nu}}\Biggr)\rightarrow \textrm{Const} ~. &&
\eeq
In practice we find that the regular solutions to Gegenbauer's equation
satisfy these boundary conditions; the singular ones do not. In fact the
regular solutions all prove to be less singular than the background and
so satisfy the more stringent boundary condition \eqref{boundary}.
The boundary conditions we impose in the other sectors are similarly motivated.

For the trivial background which doesn't have $F_{\mu\nu}$, we demand regularity at $\x=0$ and $x=\pi$.

\subsection{Charged scalar sector}

In this sector we wish to solve the scalar wave equation with minimal
coupling to the background vector potential of the monopole-antimonopole
pair. The solution closely parallels the neutral case. The natural ansatz
is to replace spherical harmonics with Wu-Yang monopole spherical harmonics:
\begin{equation}
a^{(+)}_t = \xi (\chi) W_{JM}(\theta, \phi) e^{i \omega t}.
\label{eq:chargedscalaransatz}
\end{equation}
This leads to the same equation for $\xi$ as in the neutral case,
Eq.~(\ref{eq:neutralscalarchi}),
but with the replacement $ J(J+1) \rightarrow J(J+1) - g^2$. To bring out
the parallel it is convenient to define $ \alpha $ as in
Eq.~(\ref{eq:alphadefn}) of the text, so that
$ \alpha ( \alpha + 1) = J(J+1) - g^2$ and $\alpha \geq 0$. Next we
substitute $ x = \cos \chi$ and $ \xi = (1 - x^2)^{\alpha/2} \eta$ to
find that $\eta$ obeys Gegenbauer's Eq.~(\ref{eq:gegenbauer})
with the replacement $J \rightarrow \alpha$. Solutions that are regular
at both $ x = \pm 1$ can only be found for the quantized values
$\omega^2 = (n - J + \alpha)(n - J + \alpha + 2)$. These solutions are
Gegenbauer polynomials
$G^{\alpha + \frac{1}{2}}_{n - J}$ where $n - J$ is a non-negative integer.

In summary the solutions in the charged sector are
$ \xi_{n-J, \alpha} (\chi) W_{JM}(\theta, \phi) \exp(i \omega t)$ with
frequencies $ \omega^2 = (n - J + \alpha)(n - J + \alpha + 2)$. The quantum
numbers are restricted to the range $n = g, g+1, g+2, \ldots$ and
$J = g, g+1, \ldots, n$ and $M = - J, \ldots, J$. The condition on $n$
arises from the Gegenbauer quantization condition; the limits on $J$ and
$M$ are the usual ones in the theory of monopole harmonics.

\subsection{Neutral vector sector}

We wish to determine the eigenmodes of the vector wave equation on S$^3$.
It is helpful to first solve the corresponding problem on R$^3$. It is natural
to seek solutions of the form
$f(r) {\mathbf X}^{(0)}_{JM} (\theta, \phi)$,
$f(r) {\mathbf X}^{(+)}_{JM} (\theta, \phi)$, and
$f(r) {\mathbf X}^{(-)}_{JM} (\theta, \phi)$.
The three families correspond to eigenmodes of different polarization.
Here ${\mathbf X}_{JM}(\theta, \phi)$ are the
vector spherical harmonics \cite{edmonds}. They are eigenfunctions of total angular
momentum (orbital plus spin)
with quantum numbers $J$ and $M$. They are also eigenfunctions of  orbital
angular momentum squared
with eigenvalues $J(J+1)$, $(J+1)(J+2)$ and $(J-1)J$
respectively. The quantum numbers span the range $ J = 0, 1, 2, \ldots$ and
$M = - J, \ldots, +J$. The case $J=0$ is special in that there is only one vector spherical
harmonic ${\mathbf X}^{(+)}_{00}$; the other two ${\mathbf X}^{(0)}_{00}$ and
${\mathbf X}^{(-)}_{00}$ vanish.

By use of the wave equation we determine that the radial functions $f(r)$
are spherical Bessel functions of order $J, J+1$ and $J-1$ respectively. Thus in
polar co-ordinates the solutions designated
$f(r) {\mathbf X}^{(0)}_{JM} (\theta, \phi)$ are given by
\begin{eqnarray}
v_r & = & 0
\nonumber \\
v_{\theta} & = & r j_{J}  (\omega r) \frac{1}{\sin \theta} \partial_{\phi} Y_{JM}
\nonumber \\
v_{\phi} & = & - r  j_{J} (\omega r) \sin \theta \partial_{\theta} Y_{JM};
\label{eq:canneutone}
\end{eqnarray}
the solutions $f(r) {\mathbf X}^{(+)}_{JM}$ by
\begin{eqnarray}
v_r & = & - (J+1) j_{J+1} (\omega r) Y_{JM}
\nonumber \\
v_{\theta} & = & r j_{J+1} (\omega r) \partial_{\theta} Y_{JM}
\nonumber \\
v_{\phi} & = & r j_{J+1} (\omega r) \partial_{\phi} Y_{JM};
\label{eq:canneuttwo}
\end{eqnarray}
and the solutions designated $f(r) X^{(-)}_{JM}$ by
\begin{eqnarray}
v_r & = & j_{J-1} (\omega r) J Y_{JM}
\nonumber \\
v_{\theta} & = & r j_{J-1} (\omega r) \partial_{\theta} Y_{JM}
\nonumber \\
v_{\phi} & = & r j_{J-1} (\omega r) \partial_{\phi} Y_{JM}.
\label{eq:canneutthree}
\end{eqnarray}
The frequency $\omega^2$ is continuous and restricted to the range
$\omega^2 \geq 0$.

For the subsequent generalization to S$^3$ it is necessary to
introduce a different fundamental set of
solutions. In the new set, the first family
of eigenmodes is obtained by taking the gradient of the
modes of the scalar Laplacian on R$^3$, namely $j_{J}(\omega r) Y_{JM}$.
For the second family we adopt the transverse solutions
$f(r) {\mathbf X}^{(0)}_{JM}$ given in
Eq.~(\ref{eq:canneutone}). For the third family we take the curl of the
second group of
transverse solutions. Explicitly, then the first family of solutions are
\begin{eqnarray}
v_r & = & \partial_r j_{J}(\omega r) Y_{JM}
\nonumber \\
v_{\theta} & = & j_J (\omega r) \partial_{\theta} Y_{JM}
\nonumber \\
v_{\phi} & = & j_J (\omega r) \partial_{\phi} Y_{JM}.
\label{eq:r3neutone}
\end{eqnarray}
The second family is given by Eq.~(\ref{eq:canneutone}). The third family of solutions,
obtained by taking the curl of the second family, are
\begin{eqnarray}
v_r & = & \frac{1}{r} j_J(\omega r) J (J+1) Y_{JM}
\nonumber \\
v_{\theta} & = & \partial_r [ r j_J (\omega r) ] \partial_{\theta} Y_{JM}
\nonumber \\
v_{\phi} & = & \partial_r [ r j_J (\omega r) ] \partial_{\phi} Y_{JM}.
\label{eq:r3neutthree}
\end{eqnarray}
In the special
case $J=0$ the transverse solutions and their curls vanish; only the gradient solutions
survive. This is consistent with the expectation that there is only one polarization
in this exceptional angular momentum channel.

In summary the first independent set of solutions is given by
Eqs.~(\ref{eq:canneutone}), (\ref{eq:canneuttwo}) and (\ref{eq:canneutthree});
the second set by Eqs.~(\ref{eq:r3neutone}), (\ref{eq:canneutone}) and (\ref{eq:r3neutthree}).
By making use of recursion relations for Bessel functions we can show that the
gradient solutions Eq.~(\ref{eq:r3neutone}) are the superposition
$f(r) {\mathbf X}^{(+)}_{JM} +
f(r) {\mathbf X}^{(-)}_{JM} $; the curl solutions Eq.~(\ref{eq:r3neutthree}) are the
superposition $ - J f(r) {\mathbf X}^{(+)}_{JM} + (J+1) f(r) {\mathbf X}^{(-)}_{JM}$.
Thus the two alternative sets of eigenmodes are seen to be equivalent.

We are now ready to tackle the problem on S$^3$. For the first set of eigenmodes
we try the gradient of the scalar modes leading to the ansatz
\begin{eqnarray}
a^{(3)}_{\chi} & = & \partial_{\chi} \xi_{nJ} Y_{JM}
\nonumber \\
a^{(3)}_{\theta} & = & \xi_{nJ} \partial_{\theta} Y_{JM}
\nonumber \\
a^{(3)}_{\phi} & = & \xi_{nJ} \partial_{\phi} Y_{JM}
\label{eq:s3neutone}
\end{eqnarray}
By applying the S$^3$ vector Laplacian to these modes we determine that
their eigenvalues are $n(n+2)$.

For the second set of solutions we seek purely transverse solutions. By analogy
to Eq.~(\ref{eq:canneutone}) we make the ansatz
\begin{eqnarray}
a^{(3)}_{\chi} & = & 0
\nonumber \\
a^{(3)}_{\theta} & = & f(\chi) \frac{1}{\sin \theta} \partial_{\phi} Y_{JM}
\nonumber \\
a^{(3)}_{\phi} & = & - f(\chi) \sin \theta \partial_{\theta} Y_{JM}
\label{eq:s3neuttwo}
\end{eqnarray}
Application of the S$^3$ vector Laplacian shows that these functions
are eigenmodes if we
take $f(\chi) \rightarrow \sin \chi \xi_{nJ} (\chi)$. These modes are
found to have frequency $\omega^2 = n(n+2) + 1$.

The third set of solutions are obtained by taking the curl of the
second transverse set. Again application of the S$^3$ vector Laplacian
shows that the resulting modes are
eigenfunctions with frequency $\omega^2 = n (n + 2) + 1$.

By analogy to R$^3$ we see that for the second and third set of
solutions $n = 1, 2, 3, \ldots$ and  $J = 1, 2, 3, \ldots$ and
$M = - J, \ldots, J$. For the first set, $n = 1, 2, 3, \ldots$
and $J = 0, 1, 2, \ldots$ and $M = -J, \ldots, J$.

This completes the derivation of the modes enumerated in section IV.
Note it is possible to systematically rederive these eigenmodes for
the vector Laplacian on S$^3$ using the theory of so(4) representations
just as in the scalar case. Although we will derive the relevant symmetry
generators below we will not carry out this analysis here.

\subsection{Charged vector sector}

As noted in Sec.~\ref{linearized},
the eigenmode problem in the charged vector sector
may be interpreted as the non-relativistic Schr\"{o}dinger equation
for a spin one
particle confined to a 3-sphere and moving in the magnetic
field of a monopole-antimonopole pair placed at opposite poles of the
3-sphere. The particle is minimally coupled to the magnetic field due to
it's charge as well as Zeeman coupled to the magnetic field due to its
magnetic moment.

Again, as a prelude, let us solve the simpler problem of such a particle
moving in R$^3$ in the field of a single monopole at the origin, a problem first
investigated by Brandt and Neri \cite{Brandt:1979kk}. In ref \cite{Brandt:1979kk} the focus was
on finding unstable modes by looking for imaginary eigenfrequencies. We
extend that work by enumerating all stable modes and deriving explicit
expressions for the eigenfunctions. Our primary interest in the
R$^3$ problem is that it provides an important testing ground for the
ansatze that we will later deploy on S$^3$.

In the absence of the Zeeman term
it is natural to seek three families of solutions of the form
$f(r) {\mathbf X}^{(0)}_{JM} (\theta, \phi)$,
$f(r) {\mathbf X}^{(+)}_{JM} (\theta, \phi)$, and
$f(r) {\mathbf X}^{(-)}_{JM} (\theta, \phi)$.
Here ${\mathbf X}_{JM} (\theta, \phi)$ are the monopole vector spherical harmonics
\cite{Weinberg:1993sg}. They are eigenfunctions of the total angular momentum (orbital plus
spin) with quantum numbers $J$ and $M$. They are also eigenfunctions of the total
orbital angular momentum with eigenvalue $J(J+1)$, $(J+1)(J+2)$ and $(J-1)J$
respectively. For $ g \geq 1$ the quantum numbers span the range $J= g-1, g, g+1,
\ldots$ and $M = -J, \ldots, J$. However for $J = g-1$ there is only one vector
spherical harmonic ${\mathbf X}^{(+)}$; for $J=g$ there are
two polarizations ${\mathbf X}^{(+)}$ and
${\mathbf X}^{(0)}$. For all higher values, $J = g+1, g+2, \ldots$, all three
polarizations exist. Similarly for $ g= 1/2$ the quantum numbers span the range
$ J = 1/2, 3/2, \ldots$ and $M = -J, \ldots, J$. However for $J = 1/2$ there are
only two polarizations ${\mathbf X}^{(+)}$ and ${\mathbf X}^{(0)}$. For
all higher values, $J = 3/2, 5/2, \ldots$, all three polarizations exist.

The Zeeman term couples spin and orbital motion but is invariant
under total angular momentum. By the Wigner-Eckart theorem therefore
it can only couple total angular momentum multiplets with the same quantum
numbers. Thus we conclude that for a given $J$ and $M$ there will still be
three families of solutions but they will be superpositions of the form
\begin{equation}
f(r) \left[
a \ {\mathbf X}^{(0)}_{JM} +
b \ {\mathbf X}^{(+)}_{JM} +
c \ {\mathbf X}^{(-)}_{JM} \right].
\label{eq:bnansatz}
\end{equation}
The coefficients $a, b, c$ are constants not only in that they are
independent of $(r, \theta, \phi)$ but also of $M$ (in accordance with
the Wigner-Eckart theorem). They are computed by solving for the
eigenvectors of a $3 \times 3$ matrix that is determined by substituting
the ansatz, Eq.~(\ref{eq:bnansatz}), into the eigenvalue equation for the
charged vector sector. This matrix is given explicitly by Brandt and
Neri; it can be efficiently derived making use of identities given in
Ref.~\cite{Weinberg:1993sg}.
Substitution of the ansatz, Eq.~(\ref{eq:bnansatz}), into the eigenvalue
equation also shows that $f(r)$ satisfies the radial equation for a
spinless free particle in non-relativistic quantum mechanics,
\begin{equation}
\left(
\partial_r^2 + \frac{2}{r} \partial_r  - \frac{l(l+1)}{r^2}  + \Lambda^2
\right) f = 0.
\label{eq:nrradial}
\end{equation}
with the modification that the angular momentum $l  \rightarrow \alpha - 1$
or $\alpha$ or $\alpha + 1$ corresponding to the three families of solutions.
For definiteness we shall call these families $s, t$ and $u$ respectively.
Here $\alpha$ is given by Eq.~(\ref{eq:alphadefn}) of the text.

Explicitly the solutions are given by
\begin{eqnarray}
v_r & = & [J(J+1) - g^2]  j_{\alpha - 1} W
\nonumber \\
v_{\theta} & = & (\alpha + 1) r J_{\alpha - 1} \partial_{\theta} W
\nonumber \\
v_{\phi} & = & (\alpha + 1) r j_{\alpha - 1} D_{\phi} W
\label{eq:spol}
\end{eqnarray}
for the $s$-polarization;
\begin{eqnarray}
v_r & = & 0
\nonumber \\
v_{\theta} & = & r j_{\alpha} \left(
[J(J+1) - g^2] \frac{i}{\sin \theta} D_{\phi} W + g \partial_{\theta} W \right)
\nonumber \\
v_{\phi} & = &
r j_{\alpha} \left(
g D_{\phi} W - [J(J+1) - g^2] i \sin \theta \partial_{\theta} W \right)
\nonumber \\
\label{eq:tpol}
\end{eqnarray}
for the $t$-polarization; and
\begin{eqnarray}
v_r & = & - [J(J+1) - g^2] j_{\alpha + 1} W
\nonumber \\
v_{\theta} & = &  r j_{\alpha + 1} \partial_{\theta} W
\nonumber \\
v_{\phi} & = & r j_{\alpha + 1} D_{\phi} W
\label{eq:upol}
\end{eqnarray}
for the $u$-polarization. Here $j_{\alpha}(\Lambda r)$ are spherical
Bessel functions of order $\alpha$. For brevity we have written the
Wu-Yang harmonics $W_{JM} (\theta, \phi)$ as $W$ in Eqs.~(\ref{eq:spol}), (\ref{eq:tpol}) and (\ref{eq:upol}). These modes
form a continuum with frequency $\Lambda^2 > 0$. These three families
of solutions are analogous respectively to
to Eqs.~(\ref{eq:canneutthree}), (\ref{eq:canneutone}) and (\ref{eq:canneuttwo})
to which they reduce for $g=0$. The $t$-modes derive their name from being
transverse.

There are two complications we have overlooked in Eqs.~(\ref{eq:tpol}),
(\ref{eq:spol}) and (\ref{eq:upol}). First we have ignored the special values
of $J$ for which there are fewer than three polarizations. These special cases
can be analyzed similarly. For example for case that $ J = g$
we must drop the $c$ term in the ansatz since there is no
vector spherical harmonic ${\mathbf X}^{(-)}$ in that angular momentum
channel. The upshot is that there are only two families of solutions in this
angular momentum channel. These turn out to be the $s$ and $u$ solutions
given above. It can be shown that the $t$ solution is identically zero for $J = g$.
For the case $ J = g-1$ we must drop the $b $ and $c$ in the ansatz
Eq.~(\ref{eq:bnansatz})
and there is only one family of solutions. These solutions have the form
\begin{eqnarray}
v_r & = & 0
\nonumber \\
v_{\theta} & = & r f \exp(i[M+g] \phi) (\sin \theta)^{g + M -1} (1 + \cos \theta)^{-M}
\nonumber \\
v_{\phi} & = & i r f \exp(i[M+g] \phi) (\sin \theta)^{g + M} (1 + \cos \theta)^{-M}.
\nonumber \\
\label{eq:g-1soln}
\end{eqnarray}
Here $ M = - (g-1), \ldots, (g-1)$ and $f(r)$ obeys Eq.~(\ref{eq:nrradial})
with $l(l+1) = -g$.
The angular factors in Eq.~(\ref{eq:g-1soln}) are taken from the
angular dependence of the vector spherical harmonics ${\mathbf X}^{(+)}_{g-1,M} (\theta,
\phi)$.

The second complication is that in non-relativistic quantum mechanics
$l (l+1)$ is a non-negative integer and the term involving $l$ in
Eq.~(\ref{eq:nrradial}) may be interpreted as a centrifugal barrier.
Mathematically this means there are no bound state solutions; the
continuum states are spherical Bessel functions. Here since we replace
$ l \rightarrow \alpha - 1$ or $\alpha$ or $\alpha + 1$ it is a concern
that $l(l+1)$ may become negative depending on the particular value of
$\alpha$. For the negative case Eq.~(\ref{eq:nrradial}) would have
the interpretation of a particle in an attractive inverse cube central
force potential. According to standard lore \cite{morsecube} in this
case for a weakly attractive potential, one for which
$0 > l(l+1) > - 1/4$, there are no bound states and the unbound
continuum states are still simply spherical Bessel functions.
For $ l(l+1) < -1/4$ there is a continuum of bound states with
$\Lambda^2 < 0$ as well as a continuum of unbound states with
$\Lambda^2 > 0$.

The continuum of bound states is
an artifact of treating the monopole as a point. If we worked
with a model in which the monopole had structure the singularity
in the attractive inverse square potential would be softened at
the origin. Presumably this would lead to a discrete bound state
spectrum which is generally expected for non-singular potentials.

For $J = g$, we see that $l(l+1)$ can indeed be negative for the
$s$ polarized states for $ g = 1/2, 1$ and $3/2$.
However in all these cases it is not so negative as to form bound states
(recall that for an
$s$ polarized state $l \to \alpha - 1$ where $\alpha$ is given by
Eq.~(\ref{eq:alphadefn}). It is now easy to verify that $-1/4 < l(l+1) < 0$
for $ 0 < g < 2$; otherwise $l(l+1)$ is positive).
This leads to the
important conclusion that $g = 1/2$ monopoles are stable (recall bound
states correspond to instability). For $ J = g-1$ however
$l(l+1) \rightarrow -g$
allowing the formation of bound states. Thus monopoles
with $ g \geq 1$ exhibit instability in the $J = g-1$ channel.

In summary Eqs.~(\ref{eq:spol}), (\ref{eq:tpol}) and (\ref{eq:upol})
are the eigenmodes needed to analyze the stability of a monopole
in R$^3$. For $ g = 1/2$ they are stable and represent a complete
enumeration of modes. For $g \geq 1$ there is an additional branch
of unstable modes with $J = g-1$.

The chief virtue of the derivation above is that it is systematic
and can be counted upon to yield a complete set of solutions. As in
the neutral vector sector we now rederive the eigenmodes in a second
more intuitive manner. The merit of this second derivation is
that it produces slightly simpler expressions and generalizes readily
to S$^3$. This second set of eigenmodes is also organized into three
families. The first family is obtained by starting with the gradient
solutions of the neutral vector sector, Eq.~(\ref{eq:r3neutone}) and
making the minimal coupling substitution
$\partial_{\phi} \rightarrow D_{\phi}$, replacing the spherical
harmonics with Wu-Yang harmonics and substituting
$J(J+1) \rightarrow J(J+1) - g^2$.
Finally we need to replace the Bessel functions $j_{J} (\omega r) $
with $j_{\alpha} (\omega r)$.
The third family is obtained by making the same replacements to the
curl solutions of the neutral vector sector, Eq.~(\ref{eq:r3neutthree}).
Curiously the same substitutions made to the transverse solution of the
neutral sector do not lead to a solution in the charged sector. For the
second family we therefore retain the $t$ modes of Eq.~(\ref{eq:tpol}).

We are now ready to tackle the problem on S$^3$.
For the first family of eigenmodes we start with the gradient
solution of the neutral vector sector Eq.~(\ref{eq:s3neutone}) and replace
spherical harmonics with Wu-Yang harmonics and make the minimal
coupling substitution $ \partial_{\phi} \rightarrow D_{\phi}$. This is found
to be a solution in the charged sector provided we also modify the radial
function $\xi_{n-J, J} (\chi) \rightarrow \xi_{n-J, \alpha}(\chi)$.
The eigenvalues are found to be $(n - J + \alpha)(n - J + \alpha + 2)$.
For the third family we work with the curl solution of the neutral vector
sector Eq.~(\ref{eq:neutralthree})
and in addition to the same substitutions we replace $ J(J+1) \rightarrow
J(J+1) - g^2$. This procedure too yields a solution with eigenvalue
$\Lambda^2 = (n+1 - J + \alpha)^2$. Obtaining the second family of
solutions is more challenging. We give an argument below that
there have to be transverse eigenmodes. Bolstered by this argument we make
the ansatz
\begin{eqnarray}
a^{(+)}_{\chi} & = & 0
\nonumber \\
a^{(+)}_{\theta} & = &
[ (J(J+1) - g^2 ] \frac{i}{\sin \theta} (D_{\phi} W) f +
                    g \partial_{\theta} W f
\nonumber \\
a^{(+)}_{\phi} & = & g D_{\phi} W f -
        i [J(J+1) - g^2] \sin \theta \partial_{\theta} W f
\label{eq:novansatz}
\end{eqnarray}
$f$ is a function of $ \chi$, $W$ is the Wu-Yang harmonic
$W_{JM}(\theta, \phi)$. This ansatz is motivated by the form of
transverse modes in R$^3$ and is found to yield solutions provided
$f = \sin \chi \xi_{n-J, \alpha}(\chi)$. The eigenvalues are
$( n+1 - J + \alpha)^2$.

As in R$^3$ it is possible to show that the second transverse family
does not exist for $J = g$. There is also a single branch of unstable
solutions for $J=g-1$ for $g \geq 1$. These solutions have the form
\begin{eqnarray}
a_{\chi} & = & 0
\nonumber \\
a_{\theta} & = & \sin \chi \xi \exp[i(M+g)\phi] (\sin \theta)^{g+M -1}
                         (1 + \cos \theta)^{-M}
\nonumber \\
a_{\phi} & = & i \sin \chi \xi \exp[i(M+g)\phi] (\sin \theta)^{g+M}
                           (1 + \cos \theta)^{-M}
\nonumber
\end{eqnarray}
where the radial function $\xi(\chi)$ satisfies
Eq.~(\ref{eq:neutralscalarchi}) with $J(J+1) \rightarrow -g$.
A simple analysis reveals that both solutions to this
equation diverge as $ \xi \propto 1/\sqrt{\chi}$ as
$ \chi \rightarrow 0$ and similarly as $\chi \rightarrow \pi$.
Since both solutions behave acceptably at the boundary
points there is no quantization or restriction on the frequencies
$\Lambda^2$. Hence we conclude that in the $J = g-1$ channel there
is a continuum of bound states with $\Lambda^2 < 0$ and a continuum
of unbound states with $\Lambda^2 \geq 0$.

It may seem counterintuitive that the spectrum of modes is
continuous on a finite space. Similar to the R$^3$ case, this
result is an artifact of treating the monopoles as points.
If we had worked in a model in which the monopole cores had structure
the singularity of the potential in the radial equation would be
softened at the north and south pole of S$^3$. This would lead
to a discrete spectrum, which is generally expected for non-singular
potentials. However we would still expect both bound and unbound
modes.

This concludes the derivation of the charged sector solutions and
the range of allowed quantum numbers that are given in section IV.

The main difficulty with solving Eqs.~(\ref{chargex3}), (\ref{charget3}) and
(\ref{chargep3}) is that
they are a system of coupled partial differential equations. A puzzling aspect
of our solution above is that we were able to obtain two solutions by making
simple substitutions in the neutral vector solutions but the transverse solution
did not yield to this strategy. For this reason we would like to present another
line of argument that demonstrates there must be a transverse solution. This
approach can also be applied to the neutral sector but here we concentrate
on the more vexing charged case.

We begin with the gauge condition \eqref{g+}, re-written as
\beq
    \partial_{\t}a_{\t}+\frac{\cos\t}{\sin\t}a_{\t}+
            \frac{1}{\sin^2\t}D_{\p}a_{\p} = && \nn \\
    \sin^2\x\Biggl(R_0\partial_ta_t -\partial_{\x}a_{\x}-
            2\frac{\cos\x}{\sin\x}a_{\x}  \Biggr) && ~,
\label{eq:iritgaugecondition}
\eeq
where we have suppressed the $(+)$ index. By substituting this in the $\x$ component of the vector Laplacian, Eq.~(\ref{chargex3}),
we eliminate the dependence on $a_{\t}$ and $a_{\p}$,
\beq
   \Biggl( \partial^2_{\x}+4\frac{\cos\x}{\sin\x}\partial_{\x} +
            (R \partial_t)^2+\frac{2}{\sin^2\x}-4\Biggr) a_{\x}+ && \nn \\
   \frac{1}{\sin^2\x} \Biggl( \partial_{\t}^2+
            \frac{\cos\t}{\sin\t}\partial_{\t} +
            \frac{1}{\sin^2\t}D^2_{\t} \Biggr) a_{\x} = R_0\partial_ta_t ~. &&
    \label{xt}
\eeq
Following the physical analogy to the Schroedinger equation, we now look for 3 solutions to the above equation. Repeatedly, our aim will be to somehow force separation of the $\x$ and the $(\theta, \phi)$  dependence. We will assume a trivial time dependence of the form $\exp[i\w t]$. Given that time dependence and for a general $\w$, the above equation allows one of three options - a) both $a_t$ and $a_{\x}$ are non-zero, b) $a_t$ is zero but $a_{\x}$ isn't, or c) $a_t$ and $a_{\x}$ are zero.

A reasonable first ansatz is $a_t = C a_{\x}$, $C$ being a constant.
We posit a separation of variables $a_{\x} = P(\t,\p)f(\x)$ in Eq.~\eqref{xt}, solve this
equation for $a_{\x}$, then use the resulting solution in Eqs.~(\ref{charget3}) and (\ref{chargep3})
to determine $a_{\theta}$ and $a_{\phi}$ yielding the first solution Eq.~(\ref{eq:chargedone}).

Next we attempt a solution with $a_t = 0$. Eq.~\eqref{xt} then gives a solution for
$a_{\x}$. Armed with this solution we return to Eqs.~(\ref{charget3}) and (\ref{chargep3}) and posit
that $a_{\t}$ and $a_{\p}$ have the same $\x$ dependence. To be specific we try
\beq
 a_{\x} &=&   kP(\t,\p)\frac{f(\x)}{\sin^2\x}e^{i \omega t} \nn \\
 a_{\t} &=&  \frac{G(\t,\p)}{\sin\theta} g(\x)e^{i \omega t} \nn \\
 a_{\p} &=&  i\sin\t H(\t,\p) g(\x)e^{i \omega t}
\label{eq:iritansatz2}
\eeq
where $k$ is a constant.
Plugging Eq.~(\ref{eq:iritansatz2})
into the gauge condition gives the relations that $g=\partial_{\x}f$ and
$k\sin\t P+\partial_{\t}G+iD_{\p} H = 0$. Using these relations in Eqs.~(\ref{charget3})
and (\ref{chargep3})
gives the third solution Eq.~(\ref{eq:chargethree}).

Finally let us attempt a transverse ansatz
\beq
 a_{\x} &=&   0 \nn \\
 a_{\t} &=&  \frac{G(\t,\p)}{\sin\theta} g(\x)e^{i \omega t} \nn \\
 a_{\p} &=&  i\sin\t H(\t,\p) g(\x)e^{i \omega t}  ~.
\label{eq:iritansatz3}
\eeq
Plugging this into the gauge condition, Eq.~(\ref{eq:iritgaugecondition})
gives the relation that $\partial_{\t}G=iD_{\phi}H$. We assume that $D_{\phi}H=i({\cal M} + g\A )H$.
Substitution of these results into Eqs.~(\ref{charget3}) and (\ref{chargep3}) yields
\beq
 \Biggl[\partial_x^2+\Lambda^2\Biggr] f(\x) =
        \frac{N}{\sin^2\x} f(\x) && \label{radial} \\
        \Biggl[\partial_{\t}^2+\left(\frac{\cos\t}{\sin\t}-
        2g\frac{\partial_{\t}\A}{{\cal M}+g\A}\right)\partial_{\t}  && \nn \\
 +N-\frac{({\cal M} + g\A )^2}{\sin^2\t} \Biggr] G = 0 &&  \label{angular}
\eeq
where we used $N=J(J+1)-g^2$ for brevity. While Eq.~\eqref{radial} is easily transformed into the Gegenbauer equation, the angular Eq.~\eqref{angular} is less readily solvable. Nonetheless
because the variables separate, these two equations demonstrate the consistency of the
transverse ansatz Eq.~(\ref{eq:iritansatz3}). Eq.~\eqref{angular} is essentially the Wu-Yang
monopole harmonic equation but with an extra term $2g(\partial_{\t}\A)/({\cal M}+g\A)\partial_{\t}$
that makes the equation difficult to solve. However we can {\em a posteriori}
verify that the solution Eq.~(\ref{eq:chargetwo}) is not only consistent with the transverse
ansatz Eq.~(\ref{eq:iritansatz3}) but also an explicit
solution to Eqs.~\eqref{radial} and \eqref{angular}.

\subsection{Zero-Modes}

Here we use the group theoretic technique discussed in
Appendix \ref{groupmethod} to find the translational
zero mode solutions.

First let us develop expressions for rotation generators acting on vector
fields in R$^4$. If we initially represent vectors in terms of their cartesian
components $(v_1, v_2, v_3, v_4)$ then the generator of rotations in
the $x_3$-$x_4$ plane is given by
\begin{eqnarray}
N_3 & = & i \cos \theta \partial_{\chi} -
            i \sin \theta \frac{\cos \chi}{\sin \chi} \partial_{\theta}
\nonumber \\
 & + & \left[
\begin{array}{cccc}
0 & 0 & 0 & 0 \\
0 & 0 & 0 & 0 \\
0 & 0 & 0 & -i \\
0 & 0 & i & 0
\end{array}
\right]
\label{eq:vec34}
\end{eqnarray}
This result follows from Eqs.~(\ref{eq:j3n3}) and (\ref{eq:j3n3diff}).
We would rather work with polar components which are related to cartesian
components via
\begin{equation}
\left[
\begin{array}{c}
v_1 \\
v_2 \\
v_3 \\
v_4
\end{array}
\right]
= U
\left[
\begin{array}{c}
v_r \\
v_{\chi} \\
v_{\theta} \\
v_{\phi}
\end{array}
\right]
\label{eq:cartpol}
\end{equation}
where the transformation matrix $U$ is given by
\begin{equation}
U = \left[
\begin{array}{llll}
{\rm s} \chi {\rm s} \theta {\rm c} \phi &
{\rm c} \chi {\rm s} \theta {\rm c} \phi/r &
{\rm c} \theta {\rm c} \phi / r {\rm s} \chi &
- {\rm s \phi}/r {\rm s} \chi {\rm s} \theta \\
{\rm s} \chi {\rm s} \theta {\rm s} \phi &
{\rm c} \chi {\rm s} \theta {\rm s} \phi/r &
{\rm c} \theta {\rm s} \phi/ r {\rm s} \chi &
{\rm c} \phi /r {\rm s} \chi {\rm s} \theta \\
{\rm s} \chi {\rm c} \theta &
{\rm c} \chi {\rm c} \theta/ r &
- {\rm s} \theta/r {\rm s} \chi &
0 \\
{\rm c} \chi &
- {\rm s} \chi/r &
0 &
0
\end{array}
\right]
\end{equation}
Here for brevity we have written $ \sin \theta = {\rm s} \theta$,
$\cos \chi = {\rm c} \chi$ etc. In terms of polar components $N_3$ is
given by $U^{-1} N_3 ({\rm cartesian}) U$. A simple calculation reveals the polar
form
\begin{eqnarray}
N_3 & = & i \cos \theta \partial_{\chi} - i \sin \theta
\frac{\cos \chi}{\sin \chi}
\partial_{\theta}
\nonumber \\
& + & \left[
\begin{array}{llll}
0 & 0 & 0 & 0 \\
0 & 0 & i \sin \theta/\sin^2 \chi & 0 \\
0 & - i \sin \theta & - i \cos \theta \cos \chi & 0 \\
0 & 0 & 0 & 0
\end{array}
\right].
\label{eq:vec34polar}
\end{eqnarray}
We are interested in vectors that live on the unit sphere in R$^4$. Such vectors have no
radial component and we may therefore discard the first row and column of the matrix in
Eq.~(\ref{eq:vec34polar}). The $3 \times 3$ matrix differential operators corresponding to
rotations in the other planes may be constructed similarly.

Now let us construct the zero modes. The background monopole field has polar
components $ A_{\chi} = A_{\theta} = 0$ and $A_{\phi} = g (1 - \cos \theta)$. The
zero modes are obtained by considering the change in this field under infinitesimal
rotation in the $x_1$-$x_4$, $x_2$-$x_4$ and $x_3$-$x_4$ planes, or, in other words,
by application of $N_1, N_2$ and $N_3$ to the background field configuration.
Application of $N_3$, for example, yields
$( N_3 A)_{\chi} = (N_3 A)_{\theta} = 0$ and
$( N_3 A)_{\phi} = - i g
(\cos \chi/\sin \chi) \sin^2 \theta $ which is the first zero mode listed in
section IV. The other two are obtained by application of $N_1$ and $N_2$.

\section{Group theoretic derivation of modes in neutral scalar sector}
\label{groupmethod}

It is instructive to rederive this result using the rotational symmetry
of the problem. Readers who are not interested in a group theoretic
rederivation may skip this section but should still examine
Eqs.~(\ref{eq:j3n3}) and (\ref{eq:j3n3diff}) and read the three sentences
preceding Eq.~(\ref{eq:j3n3}) and the entire paragraph containing
Eq.~(\ref{eq:j3n3diff}) as these results will be used to construct the
zero mode solutions.

Picture the sphere S$^3$ embedded in a four-dimensional space. It is
easy to write down the $ 4 \times 4$
matrices corresponding to rotation generators in the
$ x_1-x_2$, $ x_2-x_3$, $ x_3-x_1$
and $x_1-x_4$, $x_2-x_4$, $x_3-x_4$ planes. We denote these generators
$J_1, J_2, J_3$ and $N_1, N_2, N_3$ respectively. For example
\begin{equation}
J_3  =  \left[
\begin{array}{cccc}
0 & - i & 0 & 0 \\
i & 0 & 0 & 0 \\
0 & 0 & 0 & 0 \\
0 & 0 & 0 & 0
\end{array}
\right],
\hspace{2mm}
N_3 = \left[
\begin{array}{cccc}
0 & 0 & 0 & 0 \\
0 & 0 & 0 & 0 \\
0 & 0 & 0 & -i \\
0 & 0 & i & 0
\end{array}
\right]
\label{eq:j3n3}
\end{equation}
$J_3$ and $N_3$ commute and together constitute a Cartan subalgebra.
By computing the commutators of all six matrices we can obtain the
structure constants of the so(4) algebra. It is convenient to define
\begin{eqnarray}
& & K_+ = \frac{1}{2} (J_+ + N_+) ;  \hspace{2mm} M_+ = \frac{1}{2} (J_+ - N_+)
\nonumber \\
& & K_- = \frac{1}{2} (J_- + N_-);  \hspace{2mm} M_- = \frac{1}{2} (J_- - N_-)
\nonumber \\
& & K_3 = \frac{1}{2} (J_3 + N_3)  ; \hspace{2mm} M_3 = \frac{1}{2} (J_3 - N_3)
\label{eq:so4algebra}
\end{eqnarray}
where, as usual, $J_+ = J_1 + i J_2$ and $J_- = J_1 - i J_2$ and the same
for $N$. By using the explicit matrices it is easy to verify that
${\mathbf K}$ and ${\mathbf M}$ separately obey the angular momentum
algebra and commute with each other. In other words
so(4) = su(2) $\oplus$ su(2).

Working with the ${\mathbf K}, {\mathbf M}$ generators,
Eq.~(\ref{eq:so4algebra}), we see from the theory of angular momentum
that representations of the so(4) algebra can be labeled by $l_1$ and
$l_2$ which may take any value from $0, \frac{1}{2},
1, \frac{3}{2}, \ldots$ The representation $(l_1, l_2)$ has dimensionality
$ (2 l_1 + 1)(2l_2 + 1)$. The basis vectors in this representation are denoted
$| m_1, m_2 \rangle$ where $ m_1 = - l_1, \ldots, l_1$ and
$m_2 = -l_2, \ldots l_2$.
The effect of the basic algebra elements in this representation is given by
\begin{eqnarray}
K_3 | m_1, m_2 \rangle & = & m_1 | m_1, m_2 \rangle
\nonumber \\
K_+ | m_1, m_2 \rangle & = & [ (l_1 + m_1 + 1)(l_1 - m_1) ]^{\frac{1}{2}}
| m_1 + 1, m_2 \rangle
\nonumber \\
K_- | m_1, m_2 \rangle & = & [ (l_1 + m_1)(l_1 - m_1 + 1) ]^{\frac{1}{2}}
| m_1 - 1, m_2 \rangle
\nonumber \\
M_3 |m_1, m_2 \rangle & = & m_2 | m_1, m_2 \rangle
\nonumber \\
M_+ | m_1, m_2 \rangle & = & [ (l_2 + m_2 + 1)(l_2 - m_2) ]^{\frac{1}{2}}
| m_1, m_2 + 1 \rangle
\nonumber \\
M_- | m_1, m_2 \rangle & = & [ (l_2 + m_2)(l_2 - m_2 + 1) ]^{\frac{1}{2}}
| m_1, m_2 - 1 \rangle
\nonumber \\
\label{eq:l1l2rep}
\end{eqnarray}

Instead of the $|m_1, m_2 \rangle$ states it is sometimes convenient
to work with the $|JM\rangle$ basis defined by
\begin{equation}
|JM \rangle = \sum_{m_1 = - l_1}^{l_1} \sum_{m_2 = - l_2}^{l_2}
C_{m_1 m_2}^{JM} (l_1, l_2) | m_1, m_2 \rangle
\label{eq:unspeakablebasis}
\end{equation}
where $J = | l_1 - l_2 |, \ldots ,l_1 + l_2$ and $M = - J ,\ldots ,J$.
$C_{m_1 m_2}^{JM} (l_1, l_2) $ are the Clebsch-Gordan coefficients.
There are $(2 l_1 + 1)(2l_2 + 1)$ of these states, as there should be, since
the $|JM \rangle$ states are simply an alternative basis for the $(l_1, l_2)$
representation of the so(4) algebra. In this basis
the matrices for ${\mathbf J} = {\mathbf K} + {\mathbf M}$
are simple but not for ${\mathbf N}$ or ${\mathbf K}$ or ${\mathbf M}$.

The natural Casimir invariant for the so(4) algebra is
\begin{equation}
{\cal C} = {\mathbf J}^2 + {\mathbf N}^2 =
2 {\mathbf K}^2 + 2 {\mathbf M}^2
\label{eq:casimir}
\end{equation}
All the states in an $(l_1, l_2)$ representation are eigenstates of ${\cal C}$ with
eigenvalue $ 2 l_1 (l_1 + 1) + 2 l_2 (l_2 + 1) $.

Square representations where $l_1 = l_2 = l$ are of particular interest. It is helpful to define
$n = 2l$. Thus a square representations is labeled by a single integer, $n$, its order. A
square representation has dimensionality $ (2l+1)^2 = (n+1)^2$. For square representations
we see that in the  $|JM \rangle$ basis the quantum numbers span the range $ J = 0, \ldots, n$
and $ M = - J, \ldots, +J$. Moreover the Casimir invariant is given by $n(n+2)$.

Now let us consider wavefunctions $\psi( \theta, \phi, \chi)$ on S$^3$. The Hilbert space of these
wavefunctions constitutes a reducible infinite dimensional representation of the so(4) algebra.
By considering the rotation of a wavefunction we can deduce the generators of rotations in the
six fundamental planes. For example
\begin{equation}
J_3 = - i \partial_{\phi},
\hspace{2mm}
N_3 = i \cos \theta \partial_{\chi} - i \sin \theta
\frac{\cos \chi}{\sin \chi} \partial_{\theta}.
\label{eq:j3n3diff}
\end{equation}
These differential operators obey the so(4) algebra defined by their $4 \times 4$ matrix counterparts.
In deriving the differential operators it is helpful to recall that
the relation between the polar co-ordinates
of a point on S$^3$ and its four-dimensional cartesian co-ordinates is
\begin{eqnarray}
x_1 & = & \sin \chi \sin \theta \cos \phi
\nonumber \\
x_2 & = & \sin \chi \sin \theta \sin \phi
\nonumber \\
x_3 & = & \sin \chi \cos \theta
\nonumber \\
x_4 & = & \cos \chi
\label{eq:polartocartesian}
\end{eqnarray}
and that $ J_3 = - i x_1 \partial_2 + i x_2 \partial_1$,
$ N_3 = - i x_3 \partial_4 + i x_4 \partial_3$.

A simple calculation reveals that the Casimir differential operator
\begin{eqnarray}
{\cal C} & = & - \partial_{\chi}^2 - \frac{1}{\sin^2 \chi} \partial_{\theta}^2 +
\frac{1}{\sin^2 \chi \sin^2 \theta} \partial_{\phi}^2
\nonumber \\
& + & 2 \frac{\cos \chi}{\sin \chi} \partial_{\chi} +
\frac{\cos \theta}{\sin \theta}
\frac{1}{\sin^2 \chi} \partial_{\theta}.
\label{eq:diffcasimir}
\end{eqnarray}
coincides with the Laplacian on S$^3$.
Furthermore it is possible to verify that the eigenfunctions of the Laplacian
$ \xi_{nJ} (\chi) Y_{JM}(\theta, \phi)$ that we obtained by separation of
variables have the following group theoretic interpretation: For a fixed $n$,
they constitute the $|JM\rangle$ basis for a square representation of order $n$.

That only square representations are realized can be demonstrated by seeking
functions that satisfy
$K_+ \psi = 0 $ and $M_+ \psi = 0$. These are first-order equations and the solutions
are readily found to be
\begin{equation}
\psi_n = (\sin \chi)^n (\sin \theta)^n \exp(i n \phi)
\label{eq:hwsoln}
\end{equation}
where $n$ is a non-negative integer. Since these are the highest weight states of
the representation we can deduce the $(l_1, l_2)$ values of the representation by
application of $K_3$ and $M_3$ to $\psi_n$. By explicit calculation we find
$K_3 \psi_n = M_3 \psi_n = (n/2) \psi_n$ revealing that the representations
under consideration are indeed square with $l_1 = l_2 = n/2$.

\section{Bounds on $E_Q$}
\label{boundsonEQ}

In order to derive the bound first consider the contribution of the charged
sector in the absence of the monopole antimonopole
pair. The number of modes with frequency less than
or equal to $N+1$ is clearly
\begin{equation}
\Gamma_N =  \sum_{n = 1}^{N} \sum_{J = 1}^{n} \sum_{M = - J}^{J} 2
= \frac{1}{3} N (N + 1) ( 2 N + 7 ).
\label{eq:filledshell}
\end{equation}
Let us denote the frequency of the $ \Gamma^{{\rm th}}$ mode by
$\Lambda (\Gamma)$. Then $\Lambda (\Gamma )$ is
is a staircase function that jumps at $\Gamma = \Gamma_N$ where
$ N = 1, 2, \ldots$. Evidently $ \Lambda(\Gamma) = N+1$ for $\Gamma_{N-1} <
\Gamma \leq \Gamma_{N}$. In terms of this staircase function the contribution
to $ E_Q$ is given by
$ (1/2)  \int_{0}^{\Gamma_c} d \Gamma \Lambda(\Gamma)$. We can derive upper
and lower bounds by approximating the staircase $\Lambda(\Gamma)$ by
smooth functions that intersect the top of each step and the bottom of each
step respectively. Explicit calculation leads to the bounds
\begin{eqnarray}
R E_{{\rm upper}} & =  &
\frac{1}{4} \left( \frac{3}{2} \right)^{4/3} \Gamma_c^{4/3} +
\frac{1}{4} \Gamma_c
\nonumber \\
R E_{{\rm lower}} & = &
\frac{1}{4} \left( \frac{3}{2} \right)^{4/3} \Gamma_c^{4/3} -
\frac{1}{4} \Gamma_c
\label{eq:nopairbound}
\end{eqnarray}
These expressions apply in the relevant limit $ \Gamma_c \gg 1$.

The zero-point energy of the charged sector when the pair is present
can be bounded similarly. We make the additional simplification that
$\Lambda_{nJ} = n$ whilst deriving the lower bound and $\Lambda_{nJ} = n+1$
whilst deriving the upper bound. The results are:
\begin{eqnarray}
R E_{{\rm upper}} & = &
\frac{1}{4} \left( \frac{3}{2} \right)^{4/3} \Gamma_c^{4/3} +
\frac{3}{4} \Gamma_c
\nonumber \\
R E_{{\rm lower}} & = &
\frac{1}{4} \left( \frac{3}{2} \right)^{4/3} \Gamma_c^{4/3} -
\frac{1}{4} \Gamma_c
\label{eq:pairbound}
\end{eqnarray}
Upper and lower bounds on $E_Q$ may now be derived by taking the differences
of the bounds in Eqs.~(\ref{eq:nopairbound}) and (\ref{eq:pairbound}). In the discussion
so far we have concentrated upon a single charged sector. Since there are in fact
two charged sectors, $+$ and $-$, we need to double the answer, leading to the result:
\beq
- \Gamma_c < R E_Q < 2 \Gamma_c.
\label{eq:bound}
\eeq

\end{document}